\documentclass[12pt,journal,onecolumn]{IEEEtran}
\usepackage{epsfig,rotating,setspace,latexsym,amsmath,epsf,amssymb,amsfonts,bm,theorem,cite,algorithm,algorithmic,graphicx,epsf,authblk,epstopdf,url,color}

\newtheorem{theorem}{Theorem}
\newtheorem{lemma}{Lemma}
\newenvironment{Proof}[1]{\medskip\par\noindent{\bf Proof:\,}\,#1}{{\mbox{\,$\blacksquare$}\par}}

\newcommand{\bh}{{\mathbf{H}}}
\newcommand{\bg}{{\mathbf{G}}}
\newcommand{\by}{{\mathbf{Y}}}
\newcommand{\bz}{{\mathbf{Z}}}
\newcommand{\bx}{{\mathbf{X}}}
\newcommand{\bw}{{\mathbf{W}}}
\newcommand{\bs}{{\mathbf{S}}}
\newcommand{\bq}{{\mathbf{Q}}}
\newcommand{\bM}{{\mathbf{M}}}
\newcommand{\bn}{{\mathbf{N}}}
\newcommand{\bo}{{\mathbf{0}}}
\newcommand{\st}{\mathrm{s.t.}}
\newcommand{\tr}{\mathrm{tr}}

\allowdisplaybreaks

\begin{document}

\title{MIMO Wiretap Channel under Receiver Side Power Constraints with Applications to Wireless Power Transfer and Cognitive Radio}

\author{Karim Banawan \qquad Sennur Ulukus\thanks{The authors are with the Department of Electrical and Computer Engineering, University of Maryland, College Park, MD 20742 USA (e-mails: kbanawan@umd.edu; ulukus@umd.edu). This work was supported by NSF Grants CNS 13-14733, CCF 14-22111 and CCF 14-22129, and was presented in part at the Allerton conference, Monticello, IL, October 2014.}}

\renewcommand{\baselinestretch}{1.3}
\maketitle
\doublespacing

\vspace{-1.0cm}
\begin{abstract}

We consider the multiple-input multiple-output (MIMO) wiretap channel under a minimum receiver-side power constraint in addition to the usual maximum transmitter-side power constraint. This problem is motivated by energy harvesting communications with wireless energy transfer, where an added goal is to deliver a minimum amount of energy to a receiver in addition to delivering secure data to another receiver. In this paper, we characterize the exact secrecy capacity of the MIMO wiretap channel under transmitter and receiver-side power constraints. We first show that solving this problem is equivalent to solving the secrecy capacity of the wiretap channel under a \emph{double-sided correlation matrix} constraint on the channel input. We show the converse by extending the channel enhancement technique to our case. We present two achievable schemes that achieve the secrecy capacity: the first achievable scheme uses a Gaussian codebook with a fixed mean, and the second achievable scheme uses artificial noise (or cooperative jamming) together with a Gaussian codebook. The role of the mean or the artificial noise is to enable energy transfer without sacrificing from the secure rate. This is the first instance of a channel model where either the use of a mean signal or the use of channel prefixing via artificial noise is \emph{strictly necessary} for the MIMO wiretap channel. We then extend our work to consider a maximum receiver-side power constraint instead of a minimum receiver-side power constraint. This problem is motivated by cognitive radio applications, where an added goal is to decrease the received signal energy (interference temperature) at a receiver. We further extend our results to: requiring receiver-side power constraints at both receivers; considering secrecy constraints at both receivers to study broadcast channels with confidential messages; and removing the secrecy constraints to study the classical broadcast channel.
\end{abstract}

\section{Introduction}

Most existing literature on Gaussian channels is based on a transmitter-side average power constraint. This constraint models the \emph{maximum} allowable power at the transmitter-side. Gastpar \cite{Gastpar} was the first to consider a receiver-side power constraint. In \cite{Gastpar}, he considered a \emph{maximum} receiver-side power constraint motivated by the desire to limit the received interference in a cognitive radio application. He observed that, while the solution does not change with respect to a classical transmitter-side power constraint for a single-input single-output (SISO) channel, it changes significantly for a multiple-input multiple-output (MIMO) channel. Subsequently, Varshney \cite{varshney} considered a \emph{minimum} receiver-side power constraint motivated by the desire to transport both information and energy simultaneously over a wireless channel. This \emph{minimum} receiver-side power constraint signified the power (in addition to data) transferred to the receiver by the same physical signal. Varshney as well observed that while the solution does not change with respect to a classical transmitter-side power constrained SISO channel, it changes significantly with respect to a classical transmitter-side amplitude constrained SISO channel \cite{Smith71}.

In this paper, we consider a multi-user and multi-objective version of the problem considered by Gastpar and Varshney. In particular, we consider a MIMO wiretap channel where the transmitter wishes to have secure communication with a legitimate receiver in the presence of an eavesdropper. In this model, messages need to be sent at the highest reliable rate to the legitimate receiver with perfect secrecy from the eavesdropper. We impose the usual transmitter-side power constraint in addition to a receiver-side power constraint. Therefore, our model generalizes the receiver-side power constraint of Gastpar and Varshney from a single-user setting of two nodes to a multi-user scenario of a wiretap channel with three nodes, and also to a multi-objective setting where we have both reliability and security constraints.

The wiretap channel was first considered by Wyner in \cite{wyner1975wire}, where he determined the rate-equivocation region of a degraded wiretap channel. This model was generalized to arbitrary, not necessarily degraded, channels by Csiszar and Korner in \cite{csiszarKorner}, where they determined the rate-equivocation region of the most general wiretap channel. The SISO Gaussian wiretap channel, which is degraded, was considered under a transmitter-side power constraint in \cite{hellman}, which showed that Gaussian signalling is optimal. The MIMO Gaussian wiretap channel was considered in \cite{221ulukus, khistiwiretap, Hassibiwiretap}, under a transmitter-side power constraint. These references showed that channel prefixing is not needed, even though the MIMO wiretap channel is not degraded, and Gaussian signalling is optimal. An interesting alternative proof is given in \cite{chanEnhancement} based on the \emph{channel enhancement} technique developed in \cite{weingartenMIMOBC}. Reference \cite{chanEnhancement} considers the MIMO wiretap channel under a transmitter-side \emph{correlation matrix constraint} which is more general than a transmitter-side power constraint. The results in \cite{221ulukus, khistiwiretap, Hassibiwiretap, chanEnhancement} imply that artificial noise \cite{NegiAN} or cooperative jamming \cite{cooperative_jamming} is not needed for a MIMO wiretap channel under a transmitter-side power constraint.\footnote{Note, however, that they may be needed in SISO/MISO/MIMO wiretap channels with imperfect channel state information (CSI) \cite{li-yates-trappe-csi, rezki-khisti-imperfect-csi, zhou-mckay, mukherjee-swindlehurst, lin-chang-hong-etal} or  multi-user versions of the wiretap channel (e.g., multiple access) even with perfect CSI \cite{cooperative_jamming, xie-ulukus}.}

In this paper, we first characterize the secrecy capacity of the general MIMO wiretap channel under a \emph{minimum} receiver-side power constraint at the eavesdropper only. To this end, we first show that, solving the secrecy capacity of the MIMO wiretap channel under a transmitter-side \emph{maximum} power constraint and a receiver-side \emph{minimum} power constraint is equivalent to solving the secrecy capacity of a MIMO wiretap channel under a \emph{double-sided correlation matrix} constraint on the channel input at the transmitter. This is a generalization of the approach of \cite{weingartenMIMOBC, chanEnhancement}, which shows that solving the capacity under a transmitter-side \emph{maximum} power constraint is equivalent to solving the capacity under a transmitter-side \emph{maximum} correlation matrix constraint. We then generalize the channel enhancement technique of \cite{weingartenMIMOBC, chanEnhancement} to the case of \emph{double-sided} correlation matrix constraint. This gives us the converse.

We next show that the rates given in the converse can be achieved by two different achievable schemes: a \emph{mean} based scheme where the transmitter uses a Gaussian codebook with a fixed mean, and an \emph{artificial noise} \cite{NegiAN} (or cooperative jamming \cite{cooperative_jamming}) based scheme, which uses Gaussian channel prefixing with a Gaussian codebook. The role of the mean or the artificial noise is to enable energy transfer without sacrificing from the secure rate; this helps to achieve the receiver-side power constraint by sending non-message carrying signals. This is the first instance of a channel model where either the use of a mean signal or the use of channel prefixing via artificial noise is \emph{strictly necessary} for the canonical MIMO wiretap channel. Note that while \cite[Section~III]{liu-liu-poor-shamai} shows an alternative way of achieving MIMO secrecy capacity using artificial noise, this is valid in the case of a covariance constraint, and the use of artificial noise in the MIMO wiretap channel under a transmitter-side power constraint is strictly sub-optimal. We note that, in a related work, references \cite{ZhangMISOsecure, SchoberMISO} consider simultaneous information and energy transfer in a MISO wiretap channel, and focus on optimizing the performance of a specific artificial noise based achievable scheme with no claim of optimality. We also note a similar set-up in \cite{LargeSWIPTconf, LargeSWIPTjournal}, where the authors consider the case of statistical channel state information only at the transmitter and focus on optimizing asymptotic transmit covariance matrix of Gaussian codebooks without artificial noise for the case of a large number of transmit antennas.

We then extend the developed methodology to find the capacities of the following related channels. We first consider the case that both receivers (both Bob and Eve) have \emph{minimum} receiver-side power constraints. This corresponds to the case where wireless power should be delivered to both users in the system, but secure communication is guaranteed only for one of the receivers. We show that mean based or artificial noise based transmission achieves the secrecy capacity of this model. Next, we impose \emph{maximum} power constraints as opposed to \emph{minimum} power constraints at the receivers. This corresponds to a cognitive radio setting where we control the received interference power at users. In this case, we show that ordinary Gaussian signalling is sufficient, and there is no need for mean or artificial noise signalling. Next, we drop the secrecy constraint and consider the classical MIMO broadcast channel (BC) with \emph{minimum} receiver-side power constraints. This models an unsecured communication scenario where simultaneous power and information transfer is needed for both users. We prove that dirty paper coding (DPC) used in \cite{weingartenMIMOBC} is optimal to achieve the capacity. This result intuitively verifies that, even though we need \emph{minimum} received power guarantees, neither mean or artificial noise transmission is needed, because the freedom afforded by the design of the covariance matrices of the DPC scheme suffices to achieve all desired feasible receiver-side powers. Finally, we put back the secrecy constraints for both users and consider the BC with confidential messages BCCM \cite{liu-liu-poor-shamai}. We show that secure DPC (S-DPC) is optimal for the BCCM as in \cite{liu-liu-poor-shamai} without the need for mean or artificial noise signalling.

\section{System Model, Preliminaries and the Main Result} \label{system_model_wiretap}

\begin{figure}[t]
\centerline{\includegraphics[width=0.6\linewidth]{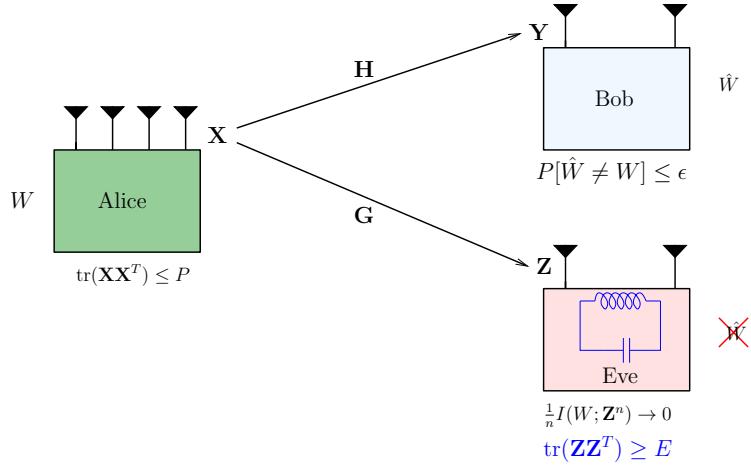}}
\caption{Gaussian MIMO wiretap channel with receiver-side power constraint.}
\label{fig_sys_model}
\vspace*{-0.5cm}
\end{figure}

The MIMO wiretap channel with $N_t$ antennas at the transmitter, $N_r$ antennas at the legitimate receiver and $N_e$ antennas at the eavesdropper is given by (see Fig.~\ref{fig_sys_model}),
\begin{align}
 \by_i&=\bh \bx_i+\bw_{1,i} \label{MIMOwiretap1} \\
 \bz_i&=\bg \bx_i+\bw_{2,i} \label{MIMOwiretap2}
\end{align}
where  $\bx_i \in \mathbb{R}^{N_t}$ is the channel input, $\by_i \in \mathbb{R}^{N_r}$ is the legitimate receiver's channel output, and $\bz_i \in \mathbb{R}^{N_e}$ is the eavesdropper's channel output at channel use $i$; $\bw_{1,i}$ and $\bw_{2,i}$ are independent Gaussian random vectors $\mathcal{N}(\bo,\mathbf{I})$. The channel matrices of legitimate receiver $\mathbf{H}$ and the eavesdropper $\mathbf{G}$ are real-valued matrices of dimensions $N_r\times N_t$ and $N_e\times N_t$, respectively, and are fixed and known to all entities. The transmitter encodes a message $W$ picked from a discrete message set $\mathcal{W}$ to a codeword $\bx^n$ over $n$ channel uses via a stochastic encoder $f:\mathcal{W} \rightarrow \bx^n$. The channel input is constrained by the usual \emph{maximum} average power constraint \cite{cover},\cite{Bloch}:
\begin{align}\label{powerconstraint}
        \frac{1}{n} \sum_{i=1}^n \tr(\bx_i\bx_i^T) \leq P
\end{align}
In this paper, we consider \emph{minimum} and \emph{maximum} power constraints at the receivers. In the initial part of the paper, we consider a \emph{minimum} power constraint at the eavesdropper only as:
\begin{align}\label{powerconstraint-eve}
       \frac{1}{n} \sum_{i=1}^n \tr(\bz_i\bz_i^T) \geq E
\end{align}
As usual, see \cite{cover, Bloch}, the actual power constraints in (\ref{powerconstraint}) and (\ref{powerconstraint-eve}) will be reflected in the single-letter capacity expressions in the sequel as expectations, i.e., $\tr(\mathbb{E}[\bx\bx^T])\leq P$ and $\tr(\mathbb{E}[\bz\bz^T])\geq E$. In addition, for all $\epsilon_n>0$, we have the following asymptotic reliability and secrecy constraints on $W$ based on $n$-length observations $\by^n,\bz^n$ at the receiver and the eavesdropper, respectively:
\begin{align} \label{confidentiality}
\mathbb{P}[\hat{W} \neq W]\leq \epsilon_n, \quad \lim_{n \to \infty} \frac{1}{n} I(W;\bz^n)=0
\end{align}
where $\epsilon_n\rightarrow 0$ as $n \rightarrow \infty$, and $\hat{W}=\phi(\by^n)$ is the estimate of the legitimate receiver of the transmitted message $W$ based on $\by^n$ by using a decoder $\phi(\cdot)$.

In this case, we have an achievable rate $R_s(E,P,\bh,\bg)=\lim_{n \rightarrow \infty}\frac{1}{n} \log |\mathcal{W}|$ if there exists a code, i.e., a codebook and $(f,\phi)$ pair such that constraints (\ref{powerconstraint})-(\ref{confidentiality}) are satisfied. The secrecy capacity $C(E,P,\bh,\bg)=\sup\: R(E,P,\bh,\bg)$, i.e., the supremum of all achievable rates. Although, we will determine the secrecy capacity under the maximum transmitter-side power constraint in (\ref{powerconstraint}) and the minimum receiver-side power constraint in (\ref{powerconstraint-eve}), we initially characterize $C(\bs_1,\bs_2,\bh,\bg)$, the secrecy capacity, under a general \emph{double-sided correlation matrix constraint}:
\begin{align}
    \bs_1\preceq \bq\preceq \bs_2 \label{corrconstraint-here}
\end{align}
where $\bq = \mathbb{E}[\bx\bx^T]$ is the channel input correlation matrix, and $\bs_1 \preceq \bs_2$ are given and fixed positive semi-definite (PSD) matrices, where $\preceq$ denotes the partial ordering of PSD matrices. We will show in a similar way to \cite[Section~II.B]{weingartenMIMOBC} that the secrecy capacity with power constraints of (\ref{powerconstraint})-(\ref{powerconstraint-eve}) can be obtained from the secrecy capacity with the more general double-sided correlation matrix constraint in (\ref{corrconstraint-here}) by maximizing this secrecy capacity over all correlation matrices $\bs_1\preceq \bs_2$ that lie in the compact set $\mathcal{S}_{PE}$:
\begin{align}
    \mathcal{S}_{PE}=\{\bs \succeq \bo:\tr(\bs)\leq P, \quad \tr(\bg\bs\bg^T)\geq \tilde{E}\} \label{SPE}
\end{align}
where $\tilde{E}=E-N_e$. We evaluate the secrecy capacity based on Csiszar-Korner secrecy capacity expression \cite{csiszarKorner}
\begin{align}\label{CsCsiszar}
        C_s=\max_{V\rightarrow \bx \rightarrow \by,\bz} I(V;\by)-I(V;\bz)
\end{align}
where $V$ carries the message signal and $\bx$ is the channel input. The maximization is over all jointly distributed $(V,\bx)$ that satisfy the Markov chain $V\rightarrow \bx \rightarrow \by,\bz$ and the constraints (\ref{powerconstraint}), (\ref{powerconstraint-eve}). Note that although Csiszar-Korner expression is initially given for discrete alphabets, it can be directly extended to alphabets other than discrete, by including the appropriate cost function in the maximization problem; see remarks in \cite[Section~VI]{csiszarKorner}. This extension can be done via discrete approximations in \cite[Chapter 3]{NWinfoTheory} and \cite[Chapter 7]{GallagerInfoTheory}.

The main result of this paper is the exact characterization of the secrecy capacity of the MIMO wiretap channel
under the maximum transmitter-side power constraint in (\ref{powerconstraint}) and the minimum receiver-side power constraint in (\ref{powerconstraint-eve}). This result is stated in Theorem~\ref{Thm1} below. We dedicate Section~\ref{sec:achievability} for the achievability proof and Section~\ref{sec:converse} for the converse proof of this theorem. In Section~\ref{sect:extensions}, we extend this basic proof technique to the cases of: minimum receiver-side power constraints at both receivers; maximum receiver-side power constraints; no secrecy constraints (classical BC); and double-sided secrecy constraints (BCCM).

\begin{theorem} \label{Thm1}
The secrecy capacity of a MIMO wiretap channel with a transmitter-side power constraint $P$ and a receiver-side power constraint $E$, $C(E,P,\bh,\bg)$, is given as
\begin{align}\label{Main result 1}
    C(E,P,\bh,\bg)=\max_{\bq \succeq \bo,\boldsymbol{\mu}} &\quad \frac{1}{2}\log |\mathbf{I}+\bh\bq\bh^T|-\frac{1}{2}\log |\mathbf{I}+\bg\bq\bg^T| \notag\\
    \st &\quad \tr({\bq+\boldsymbol{\mu}\boldsymbol{\mu}^T})\leq P, \quad
    \tr(\bg({\bq+\boldsymbol{\mu}\boldsymbol{\mu}^T})\bg^T)\geq \tilde{E}
\end{align}
where $\tilde{E}=E-N_e$. This secrecy capacity is achieved by $\bx\sim\mathcal{N}(\boldsymbol{\mu},\bq)$, i.e., with a mean but no channel prefixing. Alternatively, the secrecy capacity, $C(E,P,\bh,\bg)$, is also given as
\begin{align}\label{Main result 2}
    C(E,P,\bh,\bg)=\max_{\bq_1,\bq_2 \succeq \bo} &\quad \frac{1}{2}\log \frac{|\mathbf{I}+\bh(\bq_1+\bq_2) \bh^T|}{|\mathbf{I}+\bh\bq_2 \bh^T|}-\frac{1}{2}\log \frac{|\mathbf{I}+\bg(\bq_1+\bq_2) \bg^T|}{|\mathbf{I}+\bg\bq_2 \bg^T|} \notag\\
    \st &\quad \tr(\bq_1+\bq_2)\leq P, \quad \tr(\bg(\bq_1+\bq_2)\bg^T)\geq \tilde{E}
\end{align}
where $\bx=\mathbf{V}+\mathbf{U}$, with jointly Gaussian $\mathbf{V}\sim\mathcal{N}(\bo,\bq_1)$ and $\mathbf{U}\sim\mathcal{N}(\bo,\bq_2)$, and $\mathbf{V},\mathbf{U}$ are independent, i.e., with Gaussian signalling with Gaussian channel prefixing.
\end{theorem}

\section{Achievability Schemes} \label{sec:achievability}

In this section, we provide two coding schemes that achieve the secrecy capacity of the MIMO wiretap channel with transmitter and receiver-side power constraints given in Theorem~\ref{Thm1}.

\subsection{Gaussian Coding with Fixed Mean}

The first achievable scheme is Gaussian coding with fixed mean, i.e., $\bx \sim \mathcal{N}({\boldsymbol\mu},\bq_1)$. In this case, the fixed mean does not play a role in evaluating the secrecy capacity except for consuming part of the overall correlation matrix and only provides the required power level at the receiver side. Then, we choose $V=\bx$, i.e., no channel prefixing. Hence, we have
\begin{align}\label{fixedmean}
C(\bs_1,\bs_2,\bh,\bg) \geq \max_{\bq_1 \succeq \bo,\boldsymbol{\mu}}  & \quad I(\bx;\by)-I(\bx;\bz) \nonumber \\
= \max_{\bq_1 \succeq \bo,\boldsymbol{\mu}}  & \quad \frac{1}{2}\log |\mathbf{I}+\mathbf{H}\bq_1\bh^T|-\frac{1}{2}\log |\mathbf{I}+\bg\bq_1 \bg^T|  \nonumber\\
\st  & \quad\bs_1 \preceq \bq_1+\boldsymbol{\mu\mu}^T  \preceq \bs_2
\end{align}
In the converse proof, in place of $\boldsymbol{\mu\mu}^T$, we have a general positive semidefinite matrix $\bq_2$. In order to have a matching feasible coding scheme, $\bq_2$ must be constrained to unit-rank correlation matrices, as it corresponds to the mean of the transmitted signal. Although, the solution of $\bq_2$ is generally not unit-rank for arbitrary correlation matrices $\bs_1,\bs_2$, we show in the following lemma that for the special case of a maximum transmitter-side power constraint $P$ and a minimum receiver-side power constraint $E$, the solution is guaranteed to be of unit-rank, and hence the mean based coding scheme is feasible.

\begin{lemma} \label{lemma1-statement}
The coding scheme $\bx\sim \mathcal{N}(\mathbb{V}(\bq^*_2),\bq^*_1)$ is achievable for the wiretap channel under the transmitter-side power constraint $P$ and the receiver-side power constraint $E$ given that the matrix $\bg^T\bg$ has a unique maximum eigenvalue. The secrecy rate is characterized by the following optimization problem:
    \begin{align}\label{lemma1}
        \max_{\bq_1,\bq_2 \succeq \bo}  &\quad\frac{1}{2}\log |\mathbf{I}+\bh\bq_1\bh^T|-\frac{1}{2}\log |\mathbf{I}+\bg\bq_1 \bg^T| \notag\\
        \st &\quad \tr(\bq_1+\bq_2)\leq P, \quad \tr(\bg(\bq_1+\bq_2)\bg^T)\geq \tilde{E}
    \end{align}
where $\bq_1^*,\bq_2^*$ are the optimal correlation matrices for (\ref{lemma1}) and $\mathbb{V}(\bq^*_2)$ denotes the unique eigenvector of matrix $\bq^*_2$ with a non-zero eigenvalue.
\end{lemma}

\begin{Proof}
We note that $\bq_2$ does not appear in the objective function; it only appears in the constraint set. Therefore, its only role is to enlarge the feasible set for $\bq_1$ subject to some power constraint $\tilde{P}$, where $\tilde{P} \leq P$. Thus, $\bq_2$ must be chosen such that, when the first constraint of (\ref{lemma1}) is fixed, it maximizes the feasible set for $\bq_1$ in the second constraint, i.e., $\bq_2$ must be the solution of
\begin{align}\label{Q2_max}
        \max_{\bq_2 \succeq \bo} \quad \tr(\bg\bq_2\bg^T) \quad \st  \quad \tr(\bq_2)=\tilde{P}
\end{align}
The eigenvector decomposition for $\bq_2$, which is symmetric, is
\begin{align}\label{spectral}
  \bq_2=\sum_{i=1}^r \lambda_i \mathbf{q}_i \mathbf{q}_i^T
\end{align}
where $r$, $\lambda_i$, $\mathbf{q}_i$ are the rank, the $i$th eigenvalue and the corresponding orthonormal eigenvector of $\bq_2$, respectively. Thus, we can write the constraint as $\tr(\bq_2)=\sum_{i=1}^r \lambda_i=\tilde{P}$. Moreover, the objective function can be written as
\begin{align}\label{objective_Q2}
  \tr(\bg\bq_2\bg^T) =\tr \left(\bg\left(\sum_{i=1}^r \lambda_i \mathbf{q}_i \mathbf{q}_i^T\right)\bg^T\right)=\sum_{i=1}^r \lambda_i \|\bg\mathbf{q}_i\|^2
\end{align}
Hence, the optimization problem in (\ref{Q2_max}) can be written as
\begin{align}\label{lamda_max}
        \max_{\lambda_i,\mathbf{q}_i} \quad \sum_{i=1}^r \lambda_i \|\bg\mathbf{q}_i\|^2 \quad \st  \quad \sum_{i=1}^r \lambda_i=\tilde{P}
\end{align}
which is a linear program in $\lambda_i$. The optimum solution is $\lambda_m=\tilde{P}$, and $\lambda_i=0$ for $i \neq m$, where
\begin{align}
m = \arg \max_{i} \|\bg\mathbf{q}_i\|^2
\end{align}
Hence, the optimal solution for this problem is to beam-form all the available power $\tilde{P}$ to the direction of the largest $\|\bg\mathbf{q}_i\|^2$. This solution is unique if $\bg^T\bg$ has a unique maximum eigenvalue. Otherwise a unit-rank solution for $\bq_2$ is not guaranteed. In this case, $\bq_2=\tilde{P}\mathbf{q}_m\mathbf{q}_m^T$, i.e., it is unit-rank with eigenvector $\boldsymbol{\mu}=\sqrt{\tilde{P}}\mathbf{q}_m$, and the problem is feasible.
\end{Proof}

We remark that the same capacity expression in (\ref{lemma1}) can be realized by letting $\bx=\mathbf{V}+\mathbf{U}$, where $\mathbf{V} \sim \mathcal{N}(\bo,\bq_1)$ is the message-carrying signal and $\mathbf{U} \sim \mathcal{N}(\bo,\bq_2)$ is the energy-carrying signal that is known causally at both Bob and Eve, so that it can be cancelled prior to information decoding. We note that, with this coding scheme any covariance matrix $\bq_2$ can be realized, and therefore Lemma~1 is not needed with this coding scheme, i.e., that the converse and achievability match for all $\bs_1,\bs_2$. However, if $\bq_2$ is optimized for this scheme as well for given $P,E$, then the optimum $\bq_2$ is still unit-rank. If the problem is considered under covariance constraints, as opposed to power constraints, unit-rank requirement of the mean based scheme can be removed by sending known Gaussian signals instead, at the cost of extra overhead of identifying $\mathbf{U}$ causally at Bob and Eve.
\subsection{Gaussian Coding with Gaussian Artificial Noise}

The second achievable scheme is Gaussian coding with Gaussian artificial noise. In this case, we choose $\mathbf{X}=\mathbf{V}+\mathbf{U}$, where $\mathbf{V}$, $\mathbf{U}$ are independent and  $\mathbf{V}\thicksim\mathcal{N}(\mathbf{0},\bq_1)$ and $\mathbf{U}\thicksim\mathcal{N}(\mathbf{0}, \bq_2)$. Here, $\mathbf{V}$ carries the message, $\mathbf{X}$ is the channel input, and $\mathbf{U}$ is the artificial noise (or cooperative jamming \cite{cooperative_jamming}) signal. In this case, we use channel prefixing, hence $\mathbf{V}\neq \mathbf{X}$. The extra randomness $\mathbf{U}$ is sent by the transmitter to provide extra noise floor at both receivers, and confuses the eavesdropper. The added significance of this artificial noise in our problem is to provide a suitable level of received power at the receiver, i.e., we utilize the artificial noise as a source of power. In this case, the achievable secrecy rate satisfies
\begin{align}\label{AN}
C(\bs_1,\bs_2,\bh,\bg) \geq \max_{\bq_1,\bq_2 \succeq \bo} &\quad I(\mathbf{V};\by)-I(\mathbf{V};\bz) \nonumber \\
= \max_{\bq_1,\bq_2 \succeq \bo} &\quad \frac{1}{2}\log \frac{|\mathbf{I}+\bh(\bq_1+\bq_2) \bh^T|}{|\mathbf{I}+\bh\bq_2 \bh^T|} -\frac{1}{2}\log \frac{|\mathbf{I}+\bg(\bq_1+\bq_2) \bg^T|}{|\mathbf{I}+\bg\bq_2 \bg^T|} \nonumber\\
\st &\quad \bs_1 \preceq \bq_1+\bq_2 \preceq \bs_2
\end{align}

\section{Converse Proof} \label{sec:converse}

In this section, we  prove the reverse implication using the channel enhancement technique \cite{weingartenMIMOBC, chanEnhancement}. We will consider the case of $\bs_2\succeq \bs_1 \succ \bo$ and the aligned MIMO setting which means that the channel matrices are square and invertible. The general MIMO case follows directly from the limiting arguments in \cite{chanEnhancement}, as the additional receiver-side power constraint is irrelevant in the limit. The idea of this limiting argument is to perform singular-value decomposition of the perturbed channels $\bar{\bh}, \bar{\bg}$ \cite[Eqn.~(37)]{chanEnhancement}. Our result follows by taking the limit of this perturbation to zero. The argument is introduced in \cite[Section~II.B]{chanEnhancement} and used for example in \cite[Appendix~B.2]{liu-liu-poor-shamai}, \cite[Section~VII]{ErsenMulti}. Therefore, we focus on the aligned case here. The aligned MIMO model is obtained by multiplying the input-output relations (\ref{MIMOwiretap1})-(\ref{MIMOwiretap2}) by the inverse of the channel matrices:
\begin{align}\label{alignedMIMO}
        \tilde{\by}&=\bx+\bh^{-1}\bw_1=\bx+\tilde{\bw}_1 \\
        \tilde{\bz}&=\bx+\bg^{-1}\bw_2=\bx+\tilde{\bw}_2
\end{align}
where $\tilde{\bw}_1$ and $\tilde{\bw}_2$ are the equivalent zero-mean Gaussian random vectors with covariance matrices $\bn_1=\bh^{-1}\bh^{-T}$ and $\bn_2=\bg^{-1}\bg^{-T}$, respectively.

\subsection{Equivalence of a Double-Sided Correlation Matrix Constraint} \label{subsec:correlation}

For the MIMO broadcast and wiretap channels under a transmitter-side maximum power constraint, references \cite{weingartenMIMOBC, chanEnhancement} showed that it is sufficient to prove the converse under a maximum correlation constraint on the channel input. We first note here that in our case with maximum transmitter-side and minimum receiver-side power constraints,  a single correlation constraint on the channel input, i.e., $\bq \preceq \bs$, is not sufficient. Next, we show the equivalence of solving our problem with a \emph{double-sided} correlation matrix constraint on the channel input, i.e., $\bs_1 \preceq \bq \preceq \bs_2$. Then, our problem can be solved in two stages: the inner problem finds the capacity under fixed correlation matrices $\bs_1$ and $\bs_2$ constraints, and the outer problem finds the optimal $\bs_1, \bs_2 \in \mathcal{S}_{PE}$ in (\ref{SPE}). Finally, we modify the original channel enhancement technique \cite{weingartenMIMOBC, chanEnhancement} to prove the optimality of the achievable schemes presented in the previous section.

We first note that solving the problem for $\bq \preceq \bs$, where $\bs \in \mathcal{S}_{PE}$ is insufficient. Consider solving the secrecy capacity under maximum transmitter-side and minimum receiver-side power constraints in two stages, first, solving the problem under a fixed correlation matrix $\bs$, and then choosing the optimal $\bs\in \mathcal{S}_{PE}$, i.e.,
\begin{align}\label{singleS}
    \max_{\bs\in \mathcal{S}_{PE}}\quad \max_{\mathbf{\bq \preceq \bs}}\quad R_s(\bq,\bh,\bg)
\end{align}
where $R_s(\bq,\bh,\bg)$ is the achievable secure rate upon using correlation matrix $\bq$. Since $\bq \preceq \bs$, we have $\bg\bq\bg^T \preceq \bg\bs\bg^T$ and hence $\tr(\bg\bq\bg^T)\leq \tr(\bg\bs\bg^T)$. Then, although any $\bs \in \mathcal{S}_{PE}$ satisfies the minimum receiver-side power constraint, i.e., $\tr(\bg\bs\bg^T)\geq \tilde{E}$, the input correlation matrix $\bq$ is not guaranteed to satisfy $\tr(\bg\bq\bg^T)\geq \tilde{E}$. Hence, the single correlation constraint is not sufficient for solving problems involving minimum receiver-side power constraints.

\begin{lemma}\label{equivalence_lemma}
Since $\mathcal{S}_{PE}$ is a compact set of PSD matrices, and $C(\bs_1,\bs_2,\bh,\bg)$ is continuous with respect to $\bs_2$, we have
\begin{align}\label{equivelence}
    C(E,P,\bh,\bg)=\max_{\bs_1,\bs_2\in \mathcal{S}_{PE} ,\bs_1\preceq \bs_2} \quad C(\bs_1,\bs_2,\bh,\bg)
\end{align}
\end{lemma}

\begin{Proof}
We follow and extend the proof technique in \cite[Lemma~1]{weingartenMIMOBC} to the case of double-sided covariance matrices. We define the wiretap code $\mathcal{C}(n,\bs,R,\epsilon)$ as a codebook, where the codewords $\{\bx_i^n\}_{i=1}^{2^{nR}}$ are such that $\bs=\frac{1}{2^{nR}} \sum_{i=1}^{2^{nR}} \bx_i^n\bx_i^{n^T}$, and accompanying encoding and decoding functions $(f,\phi)$, such that $\mathbb{P}(\phi(f(W))\neq W)\leq \epsilon$. The decoder $\phi$ can be taken as the maximum likelihood decoder.

To see
\begin{align}\label{reverse}
     C(E,P,\bh,\bg)\geq \max_{\bs_1,\bs_2\in \mathcal{S}_{PE} ,\bs_1\preceq \bs_2} \quad C(\bs_1,\bs_2,\bh,\bg)
\end{align}
we note that for any $\bs_1\preceq \bq \preceq \bs_2$ where $\bs_1,\bs_2\in \mathcal{S}_{PE}$, we have $\bq\in \mathcal{S}_{PE}$, i.e., every $\bq$ in the feasible set of the optimization problem on the right hand side belongs to the feasible set of the optimization problem $C(E,P,\bh,\bg)$. Hence, $C(E,P,\bh,\bg)$ is at least as large as $\max_{\bs_1,\bs_2\in \mathcal{S}_{PE} ,\bs_1\preceq \bs_2} \: C(\bs_1,\bs_2,\bh,\bg)$.

To see
\begin{align}\label{forward}
     C(E,P,\bh,\bg)\leq \max_{\bs_1,\bs_2\in \mathcal{S}_{PE} ,\bs_1\preceq \bs_2} \quad C(\bs_1,\bs_2,\bh,\bg)
\end{align}
we should prove that $C(E,P,\bh,\bg)=C(\bs_1,\bs_2,\bh,\bg)$ for some $\bs_1,\bs_2 \in \mathcal{S}_{PE}$ \cite{weingartenMIMOBC}. If $R=C(E,P,\bh,\bg)$ is achievable, then there exists an infinite sequence of codes $\mathcal{C}(n_i,\bs_{0_i},R,\epsilon_i)$, $i=1,\dots $ with rate $R$ and decreasing probability of error $\epsilon_i\rightarrow 0$ as $i\rightarrow \infty$. Choose $\bs_1\preceq \bs_{0_i}$, $\forall i$ and $\bs_1 \in \mathcal{S}_{PE}$. We note that the choice of $\bs_1$ is completely arbitrary, thus without loss of generality, we can choose it to be the first element in the sequence, i.e., $\bs_{0_1}$. As $\mathcal{S}_{PE}$ is compact \cite{bryant1985metric, royden2010real}, for any infinite sequence of points in $\mathcal{S}_{PE}$, there must exist a sub-sequence that converges to a point $\bs_0 \in \mathcal{S}_{PE}$. Hence, for any arbitrary $\delta>0$, we can find an increasing subsequence $i(k)$ such that $\bs_1\preceq \bs_{0_{i(k)}}\preceq \bs_0 +\delta \mathbf{I} $.

This implies that we can find a sequence of codes $\mathcal{C}(n_k,\bs_0+\delta \mathbf{I},R,\epsilon_k)$ with $\bs_0 \in \mathcal{S}_{PE}$, $\bs_0\succeq\bs_1$ achieving small probability of error. Therefore, for every $\delta>0$, we have $R=C(\bs_1,\bs_0+\delta \mathbf{I},\bh,\bg)$. Since $C(\bs_1,\bs_0+\delta \mathbf{I},\bh,\bg)$ is continuous, see Appendix~\ref{continuity}, with respect to its second argument, we have that every $\epsilon$-ball around $R$ contains $C(\bs_1,\bs_0,\bh,\bg)$, since for every $\epsilon>0$, there exists $\delta>0$ such that $C(\bs_1,\bs_0+\delta \mathbf{I},\bh,\bg)-C(\bs_1,\bs_0,\bh,\bg)<\epsilon$ as continuity asserts. Therefore $R$ is a limit point of $\mathcal{C}(\bs_1,\bs_0,\bh,\bg)$ and hence $C(E,P,\bh,\bg)=C(\bs_1,\bs_0,\bh,\bg)$. This limit point belongs to $\mathcal{S}_{PE}$ since it is closed.
\end{Proof}

\subsection{Converse Proof for Gaussian Coding with Fixed Mean} \label{subsec:fixedmean}

First, we begin with writing the equivalent optimization problem corresponding to the achievability scheme in the aligned MIMO case with Gaussian coding $\bx\sim \mathcal{N}(\mathbb{V}(\bq^*_2),\bq^*_1)$:
\begin{align}\label{alignedMIMOmean}
        \max_{\bq_1,\bq_2 \succeq \bo}
        & \quad \frac{1}{2}\log \frac{|\bq_1+\bn_1|}{|\bn_1|} -\frac{1}{2}\log \frac{|\bq_1+\bn_2|}{|\bn_2|} \notag \\
        \st & \quad \bq_1+\bq_2\succeq \bs_1, \quad \bq_1+\bq_2\preceq \bs_2
\end{align}
The Lagrangian of this optimization problem can be written as:
\begin{align}\label{KKTMIMOmean}
        \mathcal{L}=&\log \frac{|\bq_1+\bn_2|}{|\bn_2|}-\log \frac{|\bq_1+\bn_1|}{|\bn_1|}
       -\tr(\bq_1\bM_1)-\tr(\bq_2\bM_2)-\tr((\bq_1+\bq_2-\bs_1)\bM_3)\notag\\
        &+\tr((\bq_1+\bq_2-\bs_2)\bM_4)
\end{align}
where $\bM_1\succeq \bo,\bM_2\succeq \bo,\bM_3\succeq \bo$ and $\bM_4\succeq \bo$ are the Lagrange multipliers for each constraint. The corresponding KKT complementary slackness conditions are:
\begin{align}
         \label{KKTmean11} \bq^*_1 \bM_1=\bo, \quad \bq^*_2 \bM_2&=\bo\\
         \label{KKTmean13}(\bq^*_1+\bq^*_2-\bs_1) \bM_3&=\bo\\
         \label{KKTmean14}(\bs_2-\bq^*_1-\bq^*_2) \bM_4&=\bo
\end{align}
and the KKT optimality conditions for $\bq^*_1$ and $\bq^*_2$ are:
\begin{align}
        (\bq^*_1+\bn_2)^{-1}-(\bq^*_1+\bn_1)^{-1}-\bM_1-\bM_3+\bM_4 &=\bo \label{KKTMIMOmean2}\\
        -\bM_2-\bM_3+\bM_4 &=\bo  \label{KKTMIMOmean3}
\end{align}
Now, using (\ref{KKTMIMOmean2}) and (\ref{KKTMIMOmean3}), we can construct an enhanced channel that can serve as an upper bound for the original legitimate receiver's channel, while the eavesdropper's channel is degraded with respect to it. The covariance of the enhanced channel is chosen as $\tilde{\bn}$ such that
\begin{align}\label{enhancedchannel}
    \begin{aligned}
        (\bq^*_1+\bn_2)^{-1}+\bM_2 =(\bq^*_1+\bn_1)^{-1}+\bM_1=(\bq^*_1+\tilde{\bn})^{-1}
    \end{aligned}
\end{align}
Using this definition of the enhanced channel, we explore various characteristics of $\tilde{\bn}$.

First, to prove the validity of the covariance matrix $\tilde{\bn}$, we note that
\begin{align}\label{Ntelde}
        \tilde{\bn}&=[(\bq^*_1+\bn_1)^{-1}+\bM_1]^{-1}-\bq^*_1\\
        &=(\mathbf{I}+\bn_1 \bM_1)^{-1} (\bq^*_1+\bn_1)-\bq^*_1\\
        &=(\mathbf{I}+\bn_1 \bM_1)^{-1} [(\bq^*_1+\bn_1)-(\mathbf{I}+\bn_1\bM_1)\bq^*_1]\\
        \label{Ntelde4}&=(\mathbf{I}+\bn_1 \bM_1)^{-1} \bn_1=(\bn_1^{-1}+\bM_1)^{-1}\succeq \bo
\end{align}
and hence the covariance matrix of the constructed enhanced channel is positive semi-definite, and therefore it is a feasible covariance matrix.

Next, we want to show that the constructed channel is enhanced with respect to $\bn_1$, i.e., $\bn_1\succeq \tilde{\bn}$. To show that we note from (\ref{Ntelde4}) that $\tilde{\bn}=(\bn_1^{-1}+\bM_1)^{-1}$ and hence, $\bn_1 \succeq \mathbf{\tilde{N}}$. Similarly by considering $(\bq^*_1+\bn_2)^{-1}+\bM_2=(\bq^*_1+\tilde{\bn})^{-1}$ we note that $\bn_2\succeq \tilde{\bn}$. Hence, we conclude that the enhanced channel has better channel conditions than the original legitimate user's channel, therefore, the constructed channel is an upper bound for the legitimate receiver. Moreover, the eavesdropper's channel is degraded with respect to the constructed channel. Consequently the secrecy capacity of the enhanced channel is known. In other words, we have $\tilde{\by}=\bx+\tilde{\bw}$ such that $\tilde{\bw} \sim \mathcal{N}(\bo,\tilde{\bn})$ and $\bx\rightarrow \tilde{\by}\rightarrow \by$ and $\bx\rightarrow \tilde{\by}\rightarrow \bz$.

In order to have a meaningful upper bound, we need to show that the rate is preserved between the original problem and the constructed channel. To show that, we have
\begin{align}\label{rate1}
        (\bq^*_1+\tilde{\bn})^{-1}\tilde{\bn}&=(\bq^*_1+\tilde{\bn})^{-1}(\tilde{\bn}+\bq^*_1-\bq^*_1)\\
        &=\mathbf{I}-(\bq^*_1+\tilde{\bn})^{-1}\bq^*_1\\
        \label{rate1a}&=\mathbf{I}-[(\bq^*_1+\bn_1)^{-1}+\bM_1]\bq^*_1\\
        \label{rate1b}&=\mathbf{I}-(\bq^*_1+\bn_1)^{-1}\bq^*_1=(\bq^*_1+\bn_1)^{-1}\bn_1
\end{align}
where (\ref{rate1a}) follows from the definition of the enhanced channel and (\ref{rate1b}) follows from the complementary slackness condition (\ref{KKTmean11}). Therefore, we have
\begin{align}\label{rate11}
        \frac{|\tilde{\bn}+\bq^*_1|}{|\tilde{\bn}|}=\frac{|\bn_1+\bq^*_1|}{|\bn_1|}
\end{align}

To show a similar rate preservation argument for the degraded channel $\bn_2$, we will need the following lemma.

\begin{lemma}\label{preservemean}
The optimal covariance matrix for the achievable scheme with Gaussian signaling with a fixed mean $\bq^*_1$ satisfies $(\bs_2-\bq^*_1)\bM_2=\bo$.
\end{lemma}

\begin{Proof}
We return to the KKT conditions. Considering the correlation constraint, three cases can possibly occur. The first case: the correlation constraint is satisfied with equality, consequently $\bs_2-\bq^*_1=\bq^*_2$. In this case, $(\bs_2-\bq^*_1)\bM_2=\bq^*_2 \bM_2=\bo$ from (\ref{KKTmean11}). The second case: the correlation constraint is strictly loose, i.e, $\bq_1+\bq_2\prec\bs_2$. In this case, we can define a matrix $\Delta=\bs_2-\bq^*_1-\bq^*_2\succ \bo$, and therefore $\Delta$ is a full-rank matrix. Thus, $\bM_4=\bo$ and from (\ref{KKTMIMOmean3}), we have $\bM_2=-\bM_3$. The matrices $\bM_2$, $\bM_3$ are both positive semi-definite matrices. Therefore, we must have $\bM_2=\bM_3=\bo$. Finally, the third case: the correlation constraint is partially loose, that is, we have $\Delta=\bs_2-\bq_1-\bq_2 \succeq \bo$, hence $\Delta$ is not a full-rank matrix. We define $\mathbf{\Sigma}=\bs_2-\bs_1 \succ \bo$, i.e., $\bs_1=\bs_2-\mathbf{\Sigma}$. In this case, we sum the KKT conditions (\ref{KKTmean13}) and (\ref{KKTmean14}) to obtain the following implications:
\begin{align}\label{delta}
    (\bq^*_1+\bq^*_2)(\bM_3-\bM_4)-\bs_1\bM_3+\bs_2\bM_4 &=\bo \\
    (\bq^*_1+\bq^*_2)(\bM_3-\bM_4)-\bs_2\bM_3+\mathbf{\Sigma}\bM_3+\bs_2\bM_4&=\bo\\
    \label{delta22}(\bs_2-\bq^*_1-\bq^*_2)(\bM_4-\bM_3)&=-\mathbf{\Sigma}\bM_3\\
    \label{delta1} (\bs_2-\bq^*_1-\bq^*_2)\bM_2&=-\mathbf{\Sigma}\bM_3\\
    \label{delta2} (\bs_2-\bq^*_1)\bM_2&=-\mathbf{\Sigma}\bM_3
\end{align}
where (\ref{delta1}) follows from (\ref{KKTMIMOmean3}), and (\ref{delta2}) follows from (\ref{KKTmean11}). Since $(\bs_2-\bq^*_1)\bM_2\succeq \bo$ and $\mathbf{\Sigma}\bM_3\succeq \bo$, or at least $(\bs_2-\bq^*_1)\bM_2$ and $\mathbf{\Sigma}\bM_3$ have the same number of non-negative eigenvalues of $\bM_2$ and $\bM_3$, respectively \cite{matrixanalysis}, the only way to satisfy (\ref{delta2}) is to have all the eigenvalues of both matrices equal zero, i.e., $(\bs_2-\bq^*_1)\bM_2=-\mathbf{\Sigma}\bM_3=\bo$. Hence, we conclude that for all three cases we have $(\bs_2-\bq^*_1)\bM_2=\bo$ and this completes the proof of Lemma~\ref{preservemean}.
\end{Proof}

Hence, using Lemma~\ref{preservemean}, we write:
\begin{align}\label{rate2}
        (\tilde{\bn}+\bs_2)(\bq^*_1+\tilde{\bn})^{-1}  &=(\bs_2-\bq^*_1)(\bq^*_1+\tilde{\bn})^{-1}+\mathbf{I}\\
        \label{rate2a} &=(\bs_2-\bq^*_1)[(\bq^*_1+\bn_2)^{-1}+\bM_2]+\mathbf{I}\\
        \label{rate2b} &=(\bs_2-\bq^*_1)(\bq^*_1+\bn_2)^{-1}+\mathbf{I}\\
        &=[(\bn_2+\bs_2)-(\bq^*_1+\bn_2)](\bq^*_1+\bn_2)^{-1}+\mathbf{I}\\
        &=(\bn_2+\bs_2)(\bq^*_1+\bn_2)^{-1}
\end{align}
where (\ref{rate2a}) follows from the definition of the enhanced channel (\ref{enhancedchannel}), and (\ref{rate2b}) follows from Lemma~\ref{preservemean}. Hence, we have:
\begin{align}\label{rate22}
    \begin{aligned}
        \frac{|\bs_2+\tilde{\bn}|}{|\bs_2+\bn_2|}=\frac{|\bq^*_1+\tilde{\bn}|}{|\bq^*_1+\bn_2|}
    \end{aligned}
\end{align}

We upper bound the secrecy capacity of the MIMO wiretap channel with a receiver-side power constraint by the secrecy capacity of the enhanced channel. Since $\bs_2 \in \mathcal{S}_{PE}$, $\bs_2$ satisfies the receiver power constraint for the enhanced channel. Hence, the receiver constraint is valid with the upper bounding argument. The secrecy capacity of the enhanced channel $\tilde{C_s}$ is given by
\begin{align}\label{Cs}
        \tilde{C_s}&=\frac{1}{2}\log\frac{|\bs_2+\tilde{\bn}|}{|\tilde{\bn}|}-\frac{1}{2}\log\frac{|\bs_2+\bn_2|}{|\bn_2|}\\
        &=\frac{1}{2}\log\frac{|\bs_2+\tilde{\bn}|}{|\bs_2+\bn_2|}\cdot\frac{|\bn_2|}{|\tilde{\bn}|}\\
        \label{Csa}&=\frac{1}{2}\log\frac{|\bq^*_1+\tilde{\bn}|}{|\bq^*_1+\bn_2|}\cdot\frac{|\bn_2|}{|\tilde{\bn}|}\\
        &=\frac{1}{2}\log\frac{|\bq^*_1+\tilde{\bn}|}{|\tilde{\bn}|}-\frac{1}{2}\log\frac{|\bq^*_1+\bn_2|}{|\bn_2|}\\
        \label{Csb} &=\frac{1}{2}\log\frac{|\bq^*_1+\bn_1|}{|\bn_1|}-\frac{1}{2}\log\frac{|\bq^*_1+\bn_2|}{|\bn_2|}=C(\bs_1,\bs_2,\bh,\bg)
\end{align}
where (\ref{Csa}) follows from (\ref{rate22}), and (\ref{Csb}) follows from (\ref{rate11}), completing the converse proof for the case of Gaussian signalling with a fixed mean.

\subsection{Converse Proof for Gaussian Coding with Gaussian Artificial Noise}

In this section, we follow a similar channel enhancement technique as in Section~\ref{subsec:fixedmean}. The optimization problem corresponding to the Gaussian coding scheme with artificial noise is:
\begin{align}\label{alignedAN}
        \max_{\bq_1,\bq_2 \succeq \bo}
        & \quad \frac{1}{2}\log \frac{|\bq_1+\bq_2+\bn_1|}{|\bq_2+\bn_1|} -\frac{1}{2}\log \frac{|\bq_1+\bq_2+\bn_2|}{|\bq_2+\bn_2|} \notag \\
        \st & \quad\bq_1+\bq_2\succeq \bs_1, \quad \bq_1+\bq_2\preceq \bs_2
\end{align}
The Lagrangian for this optimization problem is given by:
\begin{align}\label{lagrangianAN}
     \mathcal{L}=&\log \frac{|\bq_1+\bq_2+\bn_2|}{|\bq_2+\bn_2|} -\log \frac{|\bq_1+\bq_2+\bn_1|}{|\bq_2+\bn_1|}-\tr((\bq_1+\bq_2-\bs_1)\bM_3)\notag\\
        &-\tr(\bq_1 \bM_1)-\tr(\bq_2 \bM_2)+\tr((\bq_1+\bq_2-\bs_2)\bM_4)
\end{align}
The complementary slackness conditions (\ref{KKTmean11})-(\ref{KKTmean14}) are still the same due to the same set of constraints for both problems (\ref{alignedAN}) and (\ref{alignedMIMOmean}). The KKT optimality condition for $\mathbf{Q^*_1}$ and $\mathbf{Q^*_2}$ are:
\begin{align}
     \label{KKTAN2}(\bq^*_1+\bq^*_2+\bn_2)^{-1}-(\bq^*_1+\bq^*_2+\bn_1)^{-1}-\bM_1-\bM_3+\bM_4&=\bo \\
     \label{KKTAN3}
    (\bq^*_1\!+\!\bq^*_2\!+\!\bn_2)^{-1}\!-\!(\bq^*_2\!+\!\bn_2)^{-1}\!-\!(\bq^*_1\!+\!\bq^*_2\!+\!\bn_1)^{-1}\!+\!
    (\bq^*_2\!+\!\bn_1)^{-1}\!-\!\bM_2\!-\!\bM_3\!+\!\bM_4&=\bo
\end{align}
Using (\ref{KKTAN2}), we can write (\ref{KKTAN3}) as:
\begin{align}\label{KKTAN4}
      \bM_1-(\bq^*_2+\bn_2)^{-1}+(\bq^*_2+\bn_1)^{-1}-\bM_2=\bo
\end{align}
In this case, we again construct an enhanced channel with similar steps as in Section~\ref{subsec:fixedmean}. The enhanced channel is constructed as:
\begin{align}\label{enhancedchannel2}
    \begin{aligned}
        (\bq^*_2+\bn_1)^{-1}+\bM_1 =(\bq^*_2+\bn_2)^{-1}+\bM_2=(\bq_2^*+\tilde{\bn})^{-1}
    \end{aligned}
\end{align}
which is the same as in the previous section. Therefore, it follows that $\tilde{\bn}\succeq \bo$, $\tilde{\bn}\preceq \bn_1$, $\tilde{\bn}\preceq \bn_2$. Similarly, we can prove that the rate is preserved for the eavesdropper (as in the set of equations (\ref{rate1})-(\ref{rate11}) with $\bq_2^*$ instead of $\bq_1^*$), i.e.,
\begin{align}\label{rate11AN}
        \frac{|\tilde{\bn}+\bq^*_2|}{|\tilde{\bn}|}=\frac{|\bn_2+\bq^*_2|}{|\bn_2|}
\end{align}
To prove the rate preservation for the legitimate receiver, we will need the following lemma.

\begin{lemma}\label{preservAN}
To achieve a positive secrecy rate using Gaussian coding with artificial noise, $\bs_2$ must be fully used, i.e., $\bs_2=\bq^*_1+\bq^*_2$, and the optimal covariance matrix used for  the artificial noise component, $\bq^*_2$, satisfies $(\bs_2-\bq^*_2)\bM_1=\bo$.
\end{lemma}

\begin{Proof}
We start by proving the first part of the lemma by contradiction. Assume that a positive secrecy rate can be achieved using artificial noise, and $\bs_2$ is partially used. Then, we have two cases. The first case: $\Delta=\bs_2-\bq^*_1-\bq^*_2\succ \bo$. Hence, $\Delta$ is a full-rank matrix, then $\bM_4=\bo$. From (\ref{KKTAN2}), we can write $(\bq^*_1+\bq^*_2+\bn_1)^{-1}+\bM_1+\bM_3=(\bq^*_1+\bq^*_2+\bn_2)^{-1}$ and hence, $(\bq^*_1+\bq^*_2+\bn_1)^{-1}\preceq(\bq^*_1+\bq^*_2+\bn_2)^{-1}$, which results in $\bn_2\preceq \bn_1$. This means that the legitimate channel is degraded with respect to the eavesdropper channel, and hence, no positive secrecy rate can be achieved. This contradicts our assumption. The second case: $\Delta$ is not full-rank. Due to the similarity of the complementary slackness conditions for the artificial noise and the Gaussian coding with fixed mean settings, we have also (\ref{delta22}), and from (\ref{KKTAN2}), we have
\begin{align}\label{M4M3}
    &\bM_4-\bM_3=(\bq^*_1+\bq^*_2+\bn_1)^{-1}-(\bq^*_1+\bq^*_2+\bn_2)^{-1}+\bM_1
\end{align}
substituting this in (\ref{delta22}), we have the following implications:
\begin{align}\label{deltaAN}
    \Delta(\bq^*_1+\bq^*_2+\bn_1)^{-1}-\Delta(\bq^*_1+\bq^*_2+\bn_2)^{-1}+\Delta\bM_1 &=-\mathbf{\Sigma}\bM_3 \\
   \label{deltaAN3} \Delta[(\bq^*_1+\bq^*_2+\bn_2)^{-1}-\Delta(\bq^*_1+\bq^*_2+\bn_1)^{-1}]&=\Delta\bM_1+\mathbf{\Sigma}\bM_3
\end{align}
Then, $[(\bq^*_1+\bq^*_2+\bn_2)^{-1}-(\bq^*_1+\bq^*_2+\bn_1)]^{-1} \succeq \bo$ to have (\ref{deltaAN3}) hold true \cite{productsemidefinite}, and then we have $\bn_2\preceq \bn_1$ as in the previous case, which also contradicts the assumption of having a positive secrecy rate. Hence, $\bq^*_1+\bq^*_2= \bs_2$. For the second part of the lemma, we now have $\bs_2-\bq^*_2=\bq^*_1$, and from the complementary slackness condition $\bq^*_1 \bM_1=\bo$. Then, we conclude that $(\bs_2-\bq^*_2)\bM_1=\bo$, completing the proof of Lemma~\ref{preservAN}.
\end{Proof}

Using Lemma~\ref{preservAN}, we can prove rate preservation for the legitimate receiver as follows:
\begin{align}\label{rate2AN}
        (\tilde{\bn}+\bs_2)(\bq^*_2+\tilde{\bn})^{-1} &=(\bs_2-\bq^*_2)(\bq^*_2+\tilde{\bn})^{-1}+\mathbf{I}\\
        \label{rate2ANa} &=(\bs_2-\bq^*_2)[(\bq^*_2+\bn_1)^{-1}+\bM_1]+\mathbf{I}\\
        \label{rate2ANb} &=(\bs_2-\bq^*_2)(\bq^*_2+\bn_1)^{-1}+\mathbf{I}\\
        &=[(\bn_1+\bs_2)-(\bq^*_2+\bn_1)](\bq^*_2+\bn_1)^{-1}+\mathbf{I}\\
        &=(\bn_1+\bs_2)(\bq^*_2+\bn_1)^{-1}
\end{align}
where (\ref{rate2ANa}) follows from the definition of the enhanced channel (\ref{enhancedchannel2}), and (\ref{rate2ANb}) follows from Lemma~\ref{preservAN}. Therefore, we have:
\begin{align}\label{rate22AN}
    \begin{aligned}
        \frac{|\bs_2+\tilde{\bn}|}{|\bq^*_2+\tilde{\bn}|}=\frac{|\bs_2+\bn_1|}{|\bq^*_2+\bn_1|}
    \end{aligned}
\end{align}

Hence, the secrecy capacity of the enhanced channel is given by:
\begin{align}\label{CsAN}
        \tilde{C_s}&=\frac{1}{2}\log\frac{|\bs_2+\tilde{\bn}|}{|\tilde{\bn}|}-\frac{1}{2}\log\frac{|\bs_2+\bn_2|}{|\bn_2|}\\
        &=\frac{1}{2}\log\frac{|\bs_2+\tilde{\bn}|}{|\bs_2+\bn_2|}\cdot\frac{|\bn_2|}{|\tilde{\bn}|}\\
        \label{CsANa}&=\frac{1}{2}\log\frac{|\bs_2+\tilde{\bn}|}{|\bs_2+\bn_2|}\cdot\frac{|\bq^*_2+\bn_2|}{|\bq^*_2+\tilde{\bn}|}\\
        &=\frac{1}{2}\log\frac{|\bs_2+\tilde{\bn}|}{|\bq^*_2+\tilde{\bn}|}\cdot\frac{|\bq^*_2+\bn_2|}{|\bs_2+\bn_2|}\\
        \label{CsANb}&=\frac{1}{2}\log\frac{|\bs_2+\bn_1|}{|\bq^*_2+\bn_1|}\cdot\frac{|\bq^*_2+\bn_2|}{|\bs_2+\bn_2|}\\
        &=\frac{1}{2}\log \frac{|\bs_2+\bn_1|}{|\bq^*_2+\bn_1|} -\frac{1}{2}\log \frac{|\bs_2+\bn_2|}{|\bq^*_2+\bn_2|} \\
        \label{CsANc}&=\frac{1}{2}\log \frac{|\bq^*_1+\bq^*_2+\bn_1|}{|\bq^*_2+\bn_1|} -\frac{1}{2}\log \frac{|\bq^*_1+\bq^*_2+\bn_2|}{|\bq^*_2+\bn_2|}=C(\bs_1,\bs_2,\bh,\bg)
\end{align}
where (\ref{CsANa}) follows from (\ref{rate11AN}), (\ref{CsANb}) follows from (\ref{rate22AN}), and (\ref{CsANc}) follows from $\bq^*_1+\bq^*_2= \bs_2$, completing the converse proof for the case of Gaussian signalling with Gaussian artificial noise.

\section{Extensions to Related Channel Models} \label{sect:extensions}

\subsection{Gaussian MIMO Wiretap Channel Under Dual Minimum Receiver-Side Power Constraints}

In this section, we consider the case where we impose dual receiver-side \textit{minimum} power constraints, i.e., receiver-side power constraints both on the legitimate receiver and the eavesdropper. Then, we have the following constraint in addition to the constraints in (\ref{powerconstraint}) and (\ref{powerconstraint-eve}):
\begin{align}\label{Bob_constraint}
  \tr(\mathbb{E}[\by\by^T])\geq E_2
\end{align}
where $E_2$ is the minimum power level that should be delivered to the legitimate receiver. The following theorem characterizes the secrecy capacity of this model.

\begin{theorem}
The secrecy capacity of a MIMO wiretap channel with a transmitter-side power constraint $P$ and dual receiver-side power constraints $E_1, E_2$, $C(E_1,E_2,P,\bh,\bg)$, is given as
\begin{align}
    C(E_1,E_2,P,\bh,\bg) =\max_{\bq \succeq \bo,\boldsymbol{\mu}} &\quad \frac{1}{2}\log |\mathbf{I}+\bh\bq\bh^T| -\frac{1}{2}\log |\mathbf{I}+\bg\bq\bg^T| \notag\\
    \st &\quad \tr({\bq+\boldsymbol{\mu}\boldsymbol{\mu}^T})\leq P \notag\\
    &\quad \tr(\bg({\bq+\boldsymbol{\mu}\boldsymbol{\mu}^T})\bg^T)\geq \tilde{E}_1,
    \quad \tr(\bh({\bq+\boldsymbol{\mu}\boldsymbol{\mu}^T})\bh^T)\geq \tilde{E}_2
\end{align}
where $\tilde{E}_1=E_1-N_e$, and $\tilde{E}_2=E-N_r$. This secrecy capacity is achieved by $\bx\sim\mathcal{N}(\boldsymbol{\mu},\bq)$, i.e., with a mean but no channel prefixing. Alternatively, $C(E_1,E_2,P,\bh,\bg)$ is also given as
\begin{align}\label{Main result 21}
    C(E_1,E_2,P,\bh,\bg)=\max_{\bq_1,\bq_2 \succeq \bo} &\quad \frac{1}{2}\log \frac{|\mathbf{I}+\bh(\bq_1+\bq_2) \bh^T|}{|\mathbf{I}+\bh\bq_2 \bh^T|}-\frac{1}{2}\log \frac{|\mathbf{I}+\bg(\bq_1+\bq_2) \bg^T|}{|\mathbf{I}+\bg\bq_2 \bg^T|} \nonumber\\
    \st &\quad \tr(\bq_1+\bq_2)\leq P \notag\\
    &\quad \tr(\bg(\bq_1+\bq_2)\bg^T)\geq \tilde{E}_1, \quad \tr(\bh(\bq_1+\bq_2)\bh^T)\geq \tilde{E}_2
\end{align}
where $\bx=\mathbf{V}+\mathbf{U}$, with jointly Gaussian $\mathbf{V}\sim\mathcal{N}(\bo,\bq_1)$ and $\mathbf{U}\sim\mathcal{N}(\bo,\bq_2)$, where $\mathbf{U},\mathbf{V}$ are independent, i.e., Gaussian signalling with Gaussian channel prefixing.
\end{theorem}

\begin{Proof}
The proof relies on verifying that the \emph{double-sided correlation matrix constraint} constructed in Section~\ref{subsec:correlation} is sufficient for this case also. First, we define the set $\mathcal{S}_{PE_1E_2}$ as:
\begin{align}\label{corr_set_dual}
  \mathcal{S}_{PE_1E_2}=\{\bs \succeq \bo:\tr(\bs)\leq P, \quad \tr(\bg\bs\bg^T)\geq \tilde{E}_1, \quad \tr(\bh\bs\bh^T)\geq \tilde{E}_2\}
\end{align}
To show the direct implication
\begin{equation}
C(E_1,E_2,P,\bh,\bg)\geq\max_{\bs_1,\bs_2\in \mathcal{S}_{PE_1E_2} ,\bs_1\preceq \bs_2} \quad C(\bs_1,\bs_2,\bh,\bg)
\end{equation}
we note that for any $\bq$ such that $\bs_1\preceq \bq \preceq \bs_1$ where $\bs_1,\bs_2\in \mathcal{S}_{PE_1E_2}$, we have $\tr(\bq)\leq \tr(\bs_2)\leq P$, $\tr(\bg\bq\bg^T)\geq \tr(\bg\bs_1\bg^T)\geq E_1$ and $\tr(\bh\bq\bh^T)\geq \tr(\bh\bs_1\bh^T)\geq E_2$. Consequently, $\bq\in \mathcal{S}_{PE_1E_2}$, i.e., the feasible set under $\bs_1,\bs_2 \in \mathcal{S}_{PE_1E_2}$ is a subset of the feasible set under $P,E_1,E_2$ constraints. Moreover, $\mathcal{S}_{PE_1E_2}\subseteq \mathcal{S}_{PE}$ defined in Section~\ref{system_model_wiretap}, and hence $\mathcal{S}_{PE_1E_2}$ is also a compact set. Hence the implication
\begin{equation}
C(E_1,E_2,P,\bh,\bg)\leq\max_{\bs_1,\bs_2\in \mathcal{S}_{PE_1E_2} ,\bs_1\preceq \bs_2} \quad C(\bs_1,\bs_2,\bh,\bg)
\end{equation}
can be proved by following the reverse implication (\ref{reverse}) of the proof of Lemma~\ref{equivalence_lemma} for the compact set $\mathcal{S}_{PE_1E_2}$, we can show that:
\begin{align}\label{equivelence_dual}
    C(E_1,E_2,P,\bh,\bg)=\max_{\bs_1,\bs_2\in \mathcal{S}_{PE_1E_2} ,\bs_1\preceq \bs_2} \quad C(\bs_1,\bs_2,\bh,\bg)
\end{align}
Then, the inner problem under the dual receiver-side power constraints is identical to its counterpart under a single receiver-side power constraint on the eavesdropper side only. Consequently, achievability schemes of mean based and artificial noise based signalling are optimal for the dual receiver-side minimum power constraints.

It only remains to show that the achievable rates with Gaussian signalling with fixed mean match the converse, i.e., that when the covariance matrix representing the mean is left unrestricted for converse purposes, at the optimal, it takes a unit-rank so that it can be implemented with a mean vector in the achievability. That is, we need to show that Lemma~\ref{lemma1-statement} extends to the current setting under $P,E_1,E_2$ constraints. To show this, as a generalization of (\ref{Q2_max}), we need to solve:
\begin{align}\label{Q2_max_dual}
        \max_{\bq_2 \succeq \bo} \quad \alpha_1\tr(\bg\bq_2\bg^T)+\alpha_2\tr(\bh\bq_2\bh^T)\quad \st  \quad \tr(\bq_2)=\tilde{P}
\end{align}
This optimization problem is equivalent to:
\begin{align}\label{Q2_max_dual}
        \max_{\lambda_i,\mathbf{q}_i}  \quad\sum_{i=1}^r \lambda_i\left(\alpha_1\|\bg\mathbf{q}_i\|^2+\alpha_2\|\bh\mathbf{q}_i\|^2\right) \quad \st  \quad \sum_{i=1}^r \lambda_i=\tilde{P}
\end{align}
which has a beam-forming optimal solution of assigning all $\tilde{P}$ to $\mathbf{q}_m$ such that
\begin{align}
m = \arg \max_{i} \alpha_1\|\bg\mathbf{q}_i\|^2+\alpha_2\|\bh\mathbf{q}_i\|^2
\end{align}
and hence the optimal $\bq_2$ is unit-rank and the mean-based signalling is feasible.
\end{Proof}

\subsection{Gaussian MIMO Wiretap Channel Under Maximum Receiver-Side Power Constraints}

In this section, we consider the MIMO wiretap channel under \textit{maximum} receiver-side power constraints. This generalizes Gastpar's problem \cite{Gastpar} to include a secrecy requirement. In this case, we limit the interference at both receivers instead of maintaining the received power levels at both receivers as in Section~\ref{system_model_wiretap}. Then, we impose the following constraints together with (\ref{powerconstraint}):
\begin{align}\label{intereference_const}
  \tr(\mathbb{E}[\bz\bz^T])\leq E_1, \quad \tr(\mathbb{E}[\by\by^T])\leq E_2
\end{align}

\begin{theorem}
The secrecy capacity of the MIMO wiretap channel with a transmitter-side power constraint $P$ and maximum receiver-side power constraints $E_1, E_2$, $C(E_1,E_2,P,\bh,\bg)$, is
\begin{align}\label{Main result 3}
    C(E_1,E_2,P,\bh,\bg)=\max_{\bq \succeq \bo} &\quad \frac{1}{2}\log |\mathbf{I}+\bh\bq\bh^T|-\frac{1}{2}\log |\mathbf{I}+\bg\bq\bg^T| \notag\\
    \st &\quad \tr(\bq)\leq P, \quad \tr(\bg\bq\bg^T)\leq \tilde{E}_1, \quad \tr(\bh\bq\bh^T)\leq \tilde{E}_2
\end{align}
This secrecy capacity is achieved by $\bx\sim\mathcal{N}(\bo,\bq)$, i.e., neither mean or channel prefixing is required.
\end{theorem}

\begin{Proof}
Similar to the previous section, we construct a suitable correlation matrix set $\mathcal{S}_{PE_1E_2}'$ as:
\begin{align}
  \mathcal{S}_{PE_1E_2}'=\{\bs \succeq \bo:\tr(\bs)\leq P, \quad \tr(\bg\bs\bg^T)\leq \tilde{E}_1, \quad \tr(\bh\bs\bh^T)\leq \tilde{E}_2\}
\end{align}
Now, we show that, using a \emph{single-sided} correlation matrix constraint $\bq\preceq\bs$ is sufficient for \emph{maximum} receiver-side power constraints, unlike the \emph{double-sided} correlation constraint that was necessary for \emph{minimum} receiver-side power constraints so far. Since, for all $\bq \preceq \bs$, we have $\tr(\bq)\leq \tr(\bs)\leq P$, $\tr(\bg\bq\bg^T)\leq \tr(\bg\bs\bg^T)\leq \tilde{E}_1$ and $\tr(\bh\bq\bh^T)\leq \tr(\bh\bs\bh^T)\leq \tilde{E}_2$, we thus have $\bq\in\mathcal{S}_{PE_1E_2}'$. Moreover, the set $\mathcal{S}_{PE_1E_2}'$ is closed and bounded and hence compact. Consequently, we can find a sequence of codes $\mathcal{C}(n_k,\bs_0+\delta \mathbf{I},R,\epsilon_k)$ with $\bs_0 \in \mathcal{S}_{PE_1E_2}'$, achieving small probability of error, that has a limit point of $C(\bs_0,\bh,\bg)$ and hence
\begin{align}\label{equivelence_maximum}
    C(E_1,E_2,P,\bh,\bg)=\max_{\bs\in \mathcal{S}_{PE_1E_2}'} \quad C(\bs,\bh,\bg)
\end{align}
Consequently, the inner problem under a correlation matrix constraint for the wiretap channel with maximum receiver-side power limitations is identical to the inner problem for the classical wiretap channel without the extra maximum receiver-side power constraints. Hence, the classical Gaussian coding with zero-mean and no channel-prefixing is optimal.
\end{Proof}

\subsection{Gaussian MIMO Broadcast Channel Under Minimum Receiver-Side Power Constraints}

In this section, we consider the MIMO BC with no secrecy constraints under \emph{minimum} receiver-side power constraints. In this setting, the transmitter is required to communicate messages simultaneously and reliably with the largest possible rate, and at the same time, deliver the minimum required powers to the receivers: $\tr(\mathbb{E}[\bz\bz^T])\geq E_1$, $\tr(\mathbb{E}[\by\by^T])\geq E_2$. The problem without the receiver-side constraints is solved by Weingarten et. al. \cite{weingartenMIMOBC}. The rate region is achieved using DPC along with time sharing. We show in the following theorem that the DPC is optimal even after imposing the receiver-side power constraints.

\begin{theorem}
The capacity region of a MIMO broadcast channel with a transmitter-side power constraint $P$ and minimum receiver-side power constraints $E_1, E_2$, $\mathcal{C}(E_1,E_2,P,\bh,\bg)$, is given by the DPC region, which is the convex hull of the union of two regions $\mathcal{R}^{DPC}_1$ and $\mathcal{R}^{DPC}_2$, corresponding to the two orders of encoding, given as:
\begin{align}\label{DPC region}
  \mathcal{R}^{DPC}_1=&\left\{(R_1,R_2): R_1 \leq \frac{1}{2} \log |\mathbf{I}+\bh \bq_1 \bh^T|, \quad R_2 \leq \frac{1}{2} \log \frac{|\mathbf{I}+\bg(\bq_1+\bq_2)\bg^T|}{|\mathbf{I}+\bg\bq_1\bg^T|} \right\}\notag\\
  \mathcal{R}^{DPC}_2=&\left\{(R_1,R_2): R_1 \leq \frac{1}{2} \log \frac{|\mathbf{I}+\bh(\bq_1+\bq_2)\bh^T|}{|\mathbf{I}+\bh\bq_2\bh^T|}, \quad R_2 \leq \frac{1}{2} \log |\mathbf{I}+\bg \bq_2 \bg^T| \right\}
\end{align}
both of which subject to
\begin{align}
\tr(\bq_1+\bq_2)\leq P, \quad \tr(\bg(\bq_1+\bq_2)\bg^T)\geq \tilde{E}_1, \quad \tr(\bh(\bq_1+\bq_2)\bh^T)\geq \tilde{E}_2
\end{align}

\end{theorem}
\begin{Proof}
We consider, without loss of generality, the region of rates achieved by $\mathcal{R}^{DPC}_1$. We first note that, due to the presence of the minimum receiver-side power constraints, we need to consider a double-sided correlation matrix constraint $\bs_1 \preceq \bq_1+\bq_2 \preceq \bs_2$, for any fixed $\bs_1,\bs_2$ in $\mathcal{S}_{PE_1E_2}$ in (\ref{corr_set_dual}). Following the original channel enhancement proof of the aligned MIMO (not necessarily degraded) BC (AMBC) in \cite{weingartenMIMOBC}, it suffices to prove that under a double-sided correlation matrix constraint $\bs_1\preceq \bq_1+\bq_2 \preceq \bs_2$, there exists an enhanced aligned degraded BC (ADBC) such that for $\alpha_1\leq\alpha_2$, noise covariances of the enhanced channel satisfy the covariance increment $\tilde{\bn}_1 \preceq \tilde{\bn}_2$ and supporting hyperplane preservation.

First, the achievable DPC rates in the aligned case with the encoding order in $\mathcal{R}^{DPC}_1$ are
\begin{align}\label{rate_region_AMBC}
  \max_{\bq_1,\bq_2 \succeq \bo} &\quad \alpha_1\cdot\frac{1}{2}\log\frac{|\bq_1+\bn_1|}{|\bn_1|}+\alpha_2\cdot\frac{1}{2}\log\frac{|\bq_1+\bq_2+\bn_2|}{|\bq_1+\bn_2|} \notag \\
  \quad \st &\quad \bq_1+\bq_2 \succeq \bs_1, \quad \bq_1+\bq_2 \preceq \bs_2
\end{align}
The Lagrangian for this problem is:
\begin{align}\label{lagrangian_BC}
  \mathcal{L}=& \alpha_1\cdot\frac{1}{2}\log\frac{|\bq_1+\bn_1|}{|\bn_1|}+\alpha_2\cdot\frac{1}{2}\log\frac{|\bq_1+\bq_2+\bn_2|}{|\bq_1+\bn_2|}+\tr(\bq_1 \bM_1)+\tr(\bq_2 \bM_2)\notag\\
  &+\tr((\bq_1+\bq_2-\bs_1)\bM_3)-\tr((\bq_1+\bq_2-\bs_2)\bM_4)
\end{align}
The KKT optimality conditions for $\bq_1^*,\bq_2^*$ are:
\begin{align}
  \label{KKT_BC1}\frac{\alpha_1}{2}(\bq_1^*+\bn_1)^{-1}+\frac{\alpha_2}{2}(\bq_1^*+\bq_2^*+\bn_2)^{-1}-\frac{\alpha_2}{2}(\bq_1^*+\bn_2)^{-1}+\bM_1+\bM_3-\bM_4&=\bo\\
  \label{KKT_BC2}\frac{\alpha_2}{2}(\bq_1^*+\bq_2^*+\bn_2)^{-1}+\bM_2+\bM_3-\bM_4 &=\bo
\end{align}
and the complementary slackness conditions are as in (\ref{KKTmean11})-(\ref{KKTmean14}). From (\ref{KKT_BC2}) and (\ref{KKT_BC1}), we have:
\begin{align}
\label{KKTBC}\frac{\alpha_1}{2}(\bq_1^*+\bn_1)^{-1}+\bM_1=\frac{\alpha_2}{2}(\bq_1^*+\bn_2)^{-1}+\bM_2
\end{align}

Consequently, we construct the enhanced channels as:
\begin{align}
 \label{enhBC1} \frac{\alpha_1}{2}(\bq_1^*+\bn_1)^{-1}+\bM_1 &=\frac{\alpha_1}{2}(\bq_1^*+\tilde{\bn}_1)^{-1}  \\
 \label{enhBC2} \frac{\alpha_2}{2}(\bq_1^*+\bn_2)^{-1}+\bM_2 &=\frac{\alpha_2}{2}(\bq_1^*+\tilde{\bn}_2)^{-1}
\end{align}
Then, $\tilde{\bn}_1\preceq \bn_1$ and $\tilde{\bn}_2\preceq \bn_2$, and thus, the constructed channels are enhanced. We need show that the enhanced BC is degraded in favor of receiver 1. Since $\alpha_1 \leq \alpha_2$, from (\ref{KKTBC})-(\ref{enhBC2}),
\begin{align}\label{degradedADBC}
  (\bq_1^*+\tilde{\bn}_1)^{-1} = \frac{\alpha_2}{\alpha_1}(\bq_1^*+\tilde{\bn}_2)^{-1}\succeq (\bq_1^*+\tilde{\bn}_2)^{-1}
\end{align}
and hence $\tilde{\bn}_1 \preceq \tilde{\bn}_2$. Moreover, we have the rate preservation relation of receiver 1,
\begin{align}\label{rate_preserve_BC_1}
    \frac{|\bq^*_1+\tilde{\bn}_1|}{|\tilde{\bn}_1|}=\frac{|\bq^*_1+\bn_1|}{|\bn_1|}
\end{align}
and the rate preservation for user 2 can be shown as:
\begin{align}
  (\bq_1^*+\bq_2^*+\tilde{\bn}_2)(\bq_1^*+\tilde{\bn}_2)^{-1}
  &= \bq_2^*(\bq_1^*+\tilde{\bn}_2)^{-1}+\mathbf{I}  \\
  &= \bq_2^*[(\bq_1^*+\bn_2)^{-1}+\frac{2}{\alpha_2}\bM_2]+\mathbf{I}  \\
  &= \bq_2^*(\bq_1^*+\bn_2)^{-1}+\mathbf{I} \\
  &= (\bq_1^*+\bq_2^*+\bn_2)(\bq_1^*+\bn_2)^{-1}
\end{align}
leading to:
\begin{align}\label{rate_preserve_BC_2}
    \frac{|\bq^*_1+\bq^*_2+\tilde{\bn}_2|}{|\bq^*_1+\tilde{\bn}_2|}=\frac{|\bq^*_1+\bq^*_2+\bn_2|}{|\bq^*_1+\bn_2|}
\end{align}

Hence, we have an enhanced ADBC whose rate region is achieved by a Gaussian codebook and use full $\bs_2$ \cite{weingartenMIMOBC}. Additionally, from (\ref{rate_preserve_BC_1}) and (\ref{rate_preserve_BC_2}), we conclude that the rate region of the original AMBC coincides with the optimal Gaussian rate region $\mathcal{R}^G(\bs_2,\tilde{\bn}_1,\tilde{\bn}_2)$ of the enhanced ADBC. To complete the proof, we need to show that the supporting hyperplane $\{(R_1,R_2): \alpha_1 R_1 + \alpha_2 R_2=b\}$ is also a supporting hyperplane for the Gaussian rate region of the enhanced ADBC $\mathcal{R}^G(\bs_2,\tilde{\bn}_1,\tilde{\bn}_2)$, i.e., that $\sum_{i=1}^2 \alpha_i R_i^G(\bq_1,\bq_2,\tilde{\bn}_1,\tilde{\bn}_2)$ is maximized by the $\bq_i^*$ that solves the AMBC problem. The proof of this follows from \cite{weingartenMIMOBC}.
\end{Proof}

We note that the related work \cite{MISOBCzhang} considers a MISO BC with multiple receivers, where each receiver requires either data or energy, but not both. The energy-requiring users are satisfied by the transmission of pseudo-random signals, that are known to all receivers, which can be subtracted out for communication purposes with the information-requiring users. The information-requiring users are served with a DPC scheme, which is optimal in that case due to \cite{weingartenMIMOBC}, as energy transfer does not interact with data transfer. The emphasis in \cite{MISOBCzhang} is the optimization of the system for this transmission scheme. In our work, all users require both data and information simultaneously. We prove by developing a suitable channel enhancement method using double-sided correlation matrix constraints that DPC is optimal for this system.

\subsection{Gaussian MIMO Broadcast Channel with Confidential messages Under Minimum Receiver-Side Power Constraints}

In this section, we consider the MIMO BCCM where we transmit a message to each receiver secret from the other. In this setting, the transmitter is required to communicate messages reliably, securely and at the same time deliver minimum amounts of energy $E_1$ and $E_2$ to the receivers. The problem without receiver-side power constraints was solved in \cite{liu-liu-poor-shamai}, and it was shown that secure DPC (S-DPC) attains the secrecy capacity region. We show in the following theorem that S-DPC is optimal in the presence of receiver-side power constraints as well.

\begin{theorem}\label{Thm5}
The secrecy capacity region of a MIMO broadcast channel with a transmitter-side power constraint $P$ and minimum receiver-side power constraints $E_1, E_2$ and with secrecy constraints, $\mathcal{C}(E_1,E_2,P,\bh,\bg)$, is given by the S-DPC region,
\begin{align}\label{Main result 4}
    R_1 \leq \max_{\bq_1,\bq_2 \succeq \bo} &\quad \frac{1}{2}\log |\mathbf{I}+\bh\bq_1\bh^T|-\frac{1}{2}\log |\mathbf{I}+\bg\bq_1\bg^T| \notag\\
    R_2 \leq \max_{\bq_1,\bq_2 \succeq \bo} &\quad \frac{1}{2}\log \frac{|\mathbf{I}+\bg(\bq_1+\bq_2) \bg^T|}{|\mathbf{I}+\bg\bq_1 \bg^T|}-\frac{1}{2}\log \frac{|\mathbf{I}+\bh(\bq_1+\bq_2) \bh^T|}{|\mathbf{I}+\bh\bq_1 \bh^T|} \nonumber\\
    \st &\quad \tr(\bq_1+\bq_2)\leq P \nonumber\\
        &\quad\tr(\bg(\bq_1+\bq_2)\bg^T)\geq \tilde{E}_1,\quad \tr(\bh(\bq_1+\bq_2)\bh^T)\geq \tilde{E}_2
\end{align}
This region is achieved by S-DPC (Gaussian double binning) using jointly Gaussian random variables $(\mathbf{V}_1,\mathbf{V}_2)\rightarrow \bx \rightarrow (\by,\bz)$ such that $\mathbf{V}_1=\mathbf{U}_1+\mathbf{F}\mathbf{U}_2$, $\mathbf{V}_2=\mathbf{U}_2$, $\bx=\mathbf{U}_1+\mathbf{U}_2$, where $\mathbf{U}_1\sim\mathcal{N}(\bo,\bq_1)$, $\mathbf{U}_2\sim\mathcal{N}(\bo,\bq_2)$ are independent and $\mathbf{F}=\bq_1\bh^T(\mathbf{I}+\bh\bq_1\bh^T)^{-1}\bh$.
\end{theorem}

\begin{Proof}
In this case also, we have a double-sided correlation matrix constraint $\bs_1\preceq \bq_1+\bq_2\preceq\bs_2$, where $\bs_1,\bs_2$ in $\mathcal{S}_{PE_1E_2}$ in (\ref{corr_set_dual}). From Lemma~\ref{preservAN}, we know that, to have a positive secrecy rate at receiver 2, we must use the full correlation matrix $\bs_2$, i.e., $\bq_1+\bq_2=\bs_2$. Since the outer optimization problem chooses $\bs_2$ from the set $\mathcal{S}_{PE_1E_2}$, and $\bx$ has the covariance $\bq=\bq_1+\bq_2$, the receiver-side power constraints are satisfied. The achievability of the corner point follows from \cite{liu-liu-poor-shamai} by using the double binning scheme presented in \cite{liu-yates}.

We next need to show that the achievable scheme matches the converse. For receiver 2: From Theorem~\ref{Thm1}, noticing that $\bg$ in this case corresponds to the main channel and $\bh$ corresponds to the eavesdropper channel, the achievable rate $R_{2,\text{max}}$ in (\ref{Main result 4}) is equal to the secrecy capacity $C(\bs_1,\bs_2,\bg,\bh)$ in (\ref{AN}) proving the converse. For receiver 1: The achievable rate $R_{1,\text{max}}$ in (\ref{Main result 4}) is the same as the secrecy capacity $C(\bs_1,\bs_2,\bh,\bg)$ in (\ref{fixedmean}) except for the correlation constraint $\bs_1 \preceq \bq_1+\boldsymbol{\mu\mu}^T  \preceq \bs_2$. Recall that, in Section~\ref{subsec:fixedmean}, we proved the converse for arbitrary $\bq_2$, not necessarily unit-rank. Therefore, using S-DPC encoding scheme induces the required extra covariance component $\bq_2$ that supports the receiver-side constraint. Moreover, we observe that
\begin{align}\label{secrecy_capacity_equivalence}
  C(\bs_1,\bs_2,\bg,\bh)=C(\bs_1,\bs_2,\bh,\bg)+\frac{1}{2}\log \frac{|\mathbf{I}+\bg\bs_2 \bg^T|}{|\mathbf{I}+\bh\bs_2 \bh^T|}
\end{align}
This implies that $\bq_1$ maximizes the secrecy capacities of both users simultaneously. Consequently, the two users can receive the confidential messages at their respective maximum secrecy rates as individual wiretap channels, i.e., the secrecy rate region is rectangular under the $\bs_1,\bs_2$ correlation matrix constraints. Hence, the S-DPC scheme is optimal.
\end{Proof}

\section{Practical Optimization Approaches}

In this section, we provide several optimization approaches to evaluate the capacities under receiver-side power constraints stated in Theorems~\ref{Thm1}-\ref{Thm5}. Without loss of generality, we consider the case of a single minimum receiver-side power constraint in the wiretap channel in Theorem~\ref{Thm1}. This is one of the most challenging optimization problems among the results in Theorems~\ref{Thm1}-\ref{Thm5}, as the optimization problem in this case is not convex.

\subsection{MISO Problem with Gaussian Mean-Based Coding Scheme}

The MISO problem with Gaussian mean-based coding scheme can be exactly cast as a convex optimization problem by considering a linear fractional transformation (Charnes-Cooper transformation) \cite{boyd2004convex} as follows:
\begin{align}\label{MISOfixed}
        \max_{\bq_1, \bq_2 \succeq \bo}
        & \quad \frac{1}{2}\log(1+\mathbf{h}^T\bq_1\mathbf{h}) -\frac{1}{2}\log(1+\mathbf{g}^T\bq_1\mathbf{g})\notag\\
        \st &
	    \quad \tr(\bq_1)+\tr(\bq_2)\leq P, \quad
	    \mathbf{g}^T(\bq_1+\bq_2)\mathbf{g}\geq \tilde{E}
\end{align}
The objective function is generally not concave. Considering the monotonicity of $\log$, the objective function can be replaced with the linear fractional objective function         $\frac{1+\mathbf{h}^T\bq_1\mathbf{h}}{1+\mathbf{g}^T\bq_1\mathbf{g}}$.
Following the linear fractional transformation \cite{boyd2004convex} by multiplying by positive variable $t>0$ and defining $\bq_1=\tilde{\bq}_1/t$, $\bq_2=\tilde{\bq}_2/t $, and fixing the resultant denominator as $t+\mathbf{g}^T\tilde{\bq}_1\mathbf{g}=1$, we obtain the convex equivalent of the problem in (\ref{MISOfixed}) as
\begin{align}\label{MISOfixed2}
        \max_{\tilde{\bq}_1, \tilde{\bq}_2 \succeq \bo, t>0}
	    & \quad t+\mathbf{h}^T\tilde{\bq}_1\mathbf{h} \notag\\
        \st &
        \quad t+\mathbf{g}^T\tilde{\bq}_1\mathbf{g}=1\notag\\
        &\quad \tr(\tilde{\bq}_1)+\tr(\tilde{\bq}_2)\leq tP, \quad \mathbf{h}^T(\tilde{\bq}_1+\tilde{\bq}_2)\mathbf{h}\geq t\tilde{E}
\end{align}
The optimal solution of (\ref{MISOfixed2}) can be obtained efficiently using convex solvers, e.g., CVX.

\subsection{MISO Problem with Gaussian Artificial Noise Based Coding Scheme}

In this case, we cannot fully transform the problem to a convex form. However, we can apply similar techniques together with an extra step of line search \cite{linesearch} to solve the problem. The problem in this case is:
\begin{align}\label{MISOAN}
        \max_{\bq_1 , \bq_2 \succeq \bo}
        & \quad \frac{1}{2}\log \left(1+\frac{\mathbf{h}^T\bq_1\mathbf{h}}{1+\mathbf{h}^T\bq_2\mathbf{h}}\right) -\frac{1}{2}\log\underbrace{\left(1+\frac{\mathbf{g}^T\bq_1\mathbf{g}}{1+\mathbf{g}^T\bq_2\mathbf{g}}\right)}_{\leq\beta}\notag\\
        \st & \quad \tr(\bq_1)+\tr(\bq_2)\leq P,
        \quad \mathbf{g}^T(\bq_1+\bq_2)\mathbf{g}\geq \tilde{E}
\end{align}
Next, we upper bound the second term in the optimization problem by $\frac{1}{2}\log \beta$, where $\beta$ is the line-search variable. This results in an extra constraint $\frac{\mathbf{g}^T\bq_1\mathbf{g}}{1+\mathbf{g}^T\bq_2\mathbf{g}}\leq \beta-1$. We write the optimization problem by considering the monotonicity of $\log$ and rearranging terms as:
\begin{align}\label{MISOAN3}
        \max_{\bq_1 , \bq_2 \succeq \bo}
        & \quad \frac{1+\mathbf{h}^T(\bq_1+\bq_2)\mathbf{h}}{\beta (1+\mathbf{h}^T\bq_2\mathbf{h})}\notag\\
        \st& \quad \mathbf{g}^T(\bq_1-(\beta-1)\bq_2)\mathbf{g} \leq \beta-1 \notag\\
        &\quad \tr(\bq_1)+\tr(\bq_2)\leq P, \quad \mathbf{g}^T(\bq_1+\bq_2)\mathbf{g}\geq \tilde{E}
\end{align}

Now, by linear fractional transformation \cite{boyd2004convex}, we multiply (\ref{MISOAN3}) by $t>0$, define $\bq_1=\tilde{\bq}_1/t, \bq_2=\tilde{\bq}_2/t $ and fix $\beta(t+\mathbf{h}^T\tilde{\bq}_2 \mathbf{h})=1$. Note that using this transformation, the resultant problem is a convex problem for fixed $\beta$. Hence, iterating over $\beta$ along its range $1\leq\beta\leq1+P\|\mathbf{h}\|^2$, the problem becomes
\begin{align}
        \max_{\beta}
        & \quad \varphi (\beta), \quad
        \st \quad 1\leq\beta\leq1+P\|\mathbf{h}\|^2
\end{align}
which together with the following can be solved effectively
\begin{align}
        \varphi (\beta)= \max_{\tilde{\bq}_1 , \tilde{\bq}_2\succeq \bo ,t >0}
        & \quad t+\mathbf{h}^T(\tilde{\bq}_1+\tilde{\bq}_2)\mathbf{h}\notag\\
        \st &
        \quad \mathbf{g}^T(\tilde{\bq}_1-(\beta-1)\tilde{\bq}_2)\mathbf{g}\leq t(\beta-1) \notag\\
        &\quad \beta(t+\mathbf{h}^T\tilde{\bq}_2 \mathbf{h})=1\notag\\
        &\quad \tr(\mathbf{\tilde{\bq}_1})+\tr(\tilde{\bq}_2)\leq tP,
      \quad\mathbf{g}^T(\tilde{\bq}_1+\tilde{\bq}_2)\mathbf{g}\geq t\tilde{E}
\end{align}

\subsection{General MIMO Problem}

For the general MIMO case, we cannot provide a direct convex optimization equivalent as in the MISO case even by adding a line search. This is due to the concavity of log-determinant functions, which result in difference of concave functions. To tackle the problem, we can approximate the objective function using sequential convex optimization techniques \cite{SCP, convexconcave}. The idea here is to approximate the second term in the objective function by its first order expansion. To show that, first, consider the objective function of the Gaussian coding with fixed mean $\frac{1}{2}\log |\mathbf{I}+\bh\bq_1\bh^T| -\frac{1}{2}\log |\mathbf{I}+\bg\bq_1\bg^T|$, which is equivalent to $\log |\bq_1+\bn_1|-\log |\bq_1+\bn_2|$. We approximate the second term with an affine function using the Taylor series expansion of the $\log \det$ function around $\bq^{(k)}$, where $k$ denotes the $k$th iteration:
\begin{align}\label{Taylor}
        \log |\bq_1+\bn_2|\cong \log |\bq^{(k)}_1+\bn_2|+\tr((\bq^{(k)}_1+\bn_2)^{-1}(\bq_1-\bq^{(k)}))
\end{align}
Since the constant terms do not affect the optimal solution, we can use
\begin{align}\label{Taylor2}
    \log |\bq_1+\bn_2| \cong \tr((\bq^{(k)}_1+\bn_2)^{-1}\bq_1)
\end{align}
The optimization problem in the $k$th iteration is
\begin{align}\label{MIMOfixed}
        \max_{\bq_1 , \bq_2 \succeq \bo}
        & \quad \log |\bq_1+\bn_1|-\tr((\bq^{(k)}_1+\bn_2)^{-1}\bq_1)\notag\\
        \st &\quad \tr(\bq_1)+\tr(\bq_2)\leq P, \quad \tr(\bg(\bq_1+\bq_2)\bg^T)\geq \tilde{E}
\end{align}
which is a convex problem, and can be solved efficiently. We update $\bq^{(k)}_1,\bq^{(k)}_2$ by solving such convex optimization problems until convergence.

Finally, using similar ideas, we can perform linearization in the case of Gaussian with artificial noise coding scheme, where the corresponding optimization problem in the $k$th iteration is
\begin{align}\label{MIMOAN}
        \max_{\bq_2,\bs \succeq \bo}
        & \quad \log |\bs+\bn_1|+\log |\bq_2+\bn_2|-\tr((\bq_2^{(k)}+\bn_1)^{-1}\bq_2)-\tr((\bs^{(k)}+\bn_2)^{-1}\bs)\notag\\
        \st & \quad\tr(\bs)\leq P, \quad\tr(\bg\bs\bg^T)\geq \tilde{E}
\end{align}

\section{Numerical Results}
In this section, we present simple simulation results for the secrecy capacity of the MIMO wiretap channel with maximum transmitter-side power constraint and minimum receiver-side (eavesdropper-side) power constraint. In these simulations, the average transmit power at the transmitter is taken as $P=10$ and the noise covariance is identity at both receivers.

\begin{figure}[t]
\centerline{\includegraphics[width=0.62\linewidth]{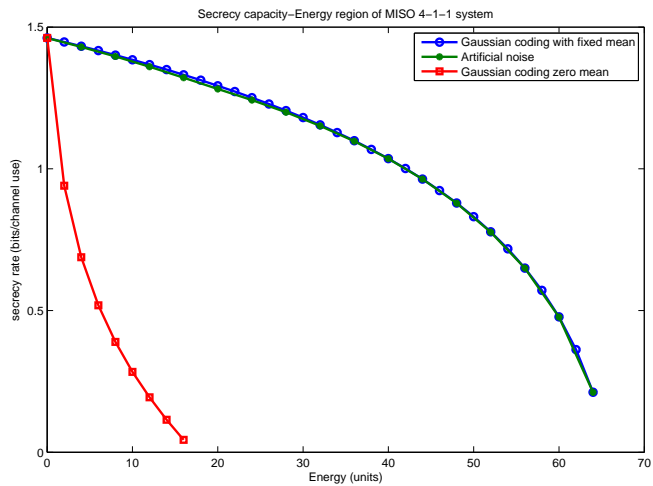}}
\caption{Secrecy capacity receiver-side power constraint region for a 4-1-1 MISO wiretap channel.}
\label{fig_MISO}
\vspace*{-0.3cm}
\end{figure}

\begin{figure}[t]
\centerline{\includegraphics[width=0.62\linewidth]{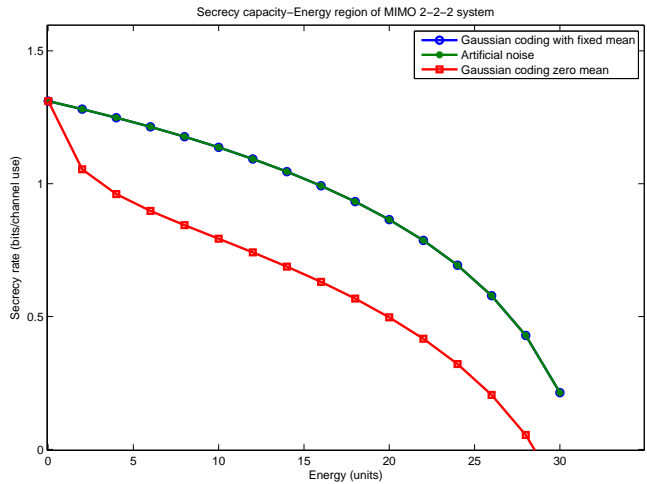}}
\caption{Secrecy capacity receiver-side power constraint region for a 2-2-2 MIMO wiretap channel.}
\label{fig_MIMO}
\vspace*{-0.3cm}
\end{figure}

Fig.~\ref{fig_MISO} shows a secrecy capacity receiver-side power constraint region for a MISO 4-1-1 system, i.e, a system with 4 antennas at the transmitter and single antenna at both the legitimate receiver and the eavesdropper. The figure shows the optimality of the Gaussian signalling with a mean and Gaussian coding with Gaussian artificial noise coding schemes; in particular, the regions corresponding to the mean and artificial noise coding schemes are identical. Moreover, the secrecy rate region with receiver-side power region of the standard Gaussian coding scheme with no mean or no artificial noise is noticeably smaller than the optimal schemes. That is, the standard Gaussian signaling scheme is strictly sub-optimal for the case of receiver-side power constraints. In addition, we observe that, as the receiver-side power constraint is increased, the secrecy capacity decreases, i.e., there is a trade-off between the power that should be delivered to the eavesdropper's receiver and the confidentiality that can be provided to the legitimate receiver. This is because, when the receiver-side power constraint is increased, the problem becomes more confined and more power should be concentrated for the receiver-side power constraint, which decreases the set of signalling choices for the secrecy communications. Fig.~\ref{fig_MIMO} shows similar observations for the 2-2-2 MIMO wiretap system.

\section{Conclusions}

We considered the MIMO wiretap channel with the usual transmitter-side maximum power constraint and an additional receiver-side minimum power constraint. For the converse, we first proved that the problem is equivalent to solving a secrecy capacity problem with a double-sided correlation matrix constraint on the channel input. We then extended the channel enhancement technique to our setting.  For the achievability, we proposed two optimum schemes that achieve the converse rate: Gaussian signalling with a fixed mean and Gaussian signalling with Gaussian channel prefixing (artificial noise). This is the first instance of a problem where transmission with a mean or channel prefixing are strictly necessary for a MIMO wiretap channel under power constraints. The transmission scheme with a mean enables us to deliver the needed power to the receiver without creating interference to the legitimate receiver as it is a deterministic signal. On the other hand, the transmission scheme with Gaussian artificial noise, both jams the eavesdropper contributing to the secrecy as well as delivering the needed power to the receiver. We note that the optimal coding scheme for the MIMO wiretap channel under a transmitter-side power constraint only, which is Gaussian signalling with no channel prefixing or mean, is strictly sub-optimal when we impose a receiver-side power constraint, showing similar to the cases of \cite{Gastpar, varshney}, that receiver-side power constraints may change the solution significantly and may introduce non-trivial trade-offs. We then extended our setting to the cases of minimum power constraints at both receivers in a wiretap channel; maximum receiver-side power constraints at both receivers in a wiretap channel; minimum receiver-side power constraints in a broadcast channel (i.e., no secrecy constraints); and minimum receiver-side power constraints in a broadcast channel with confidential messages (i.e., double-sided secrecy constraints).

\appendix[Continuity of the Capacity Function]\label{continuity}
We prove our claim in Lemma~\ref{equivalence_lemma} that $C(\bs_1,\bs_2,\bh,\bg)$ is a continuous function with respect to $\bs_2$. Although contiguity defined in \cite{weingartenMIMOBC}, which is a weaker notion than continuity, suffices to prove Lemma~\ref{equivalence_lemma}, we prove continuity here. To prove this,  we begin by writing the optimization problem in a general form as in \cite[Appendix IV]{weingartenMIMOBC} by concatenating the rows of $\bq_1,\bq_2$ to form a vector $\mathbf{y} \in \mathbb{R}^{2t^2}$, where $t=\max\{N_t,N_r\}$. We denote the point-to-set map $\Omega(\bs_2)$ to be a mapping from $\bs_2$ to the power set of all subsets of the corresponding  feasible set, i.e.,
\begin{equation}
\Omega(\bs_2)=\{\text{row concatenation of }(\bq_1,\bq_2): \bq_1,\bq_2 \succeq \bo,\: \bs_1 \preceq \bq_1+\bq_2 \preceq \bs_2\}
\end{equation}
Denote $C(\bs_1,\bs_2,\bh,\bg)$ by $C(\bs_2)$ for notational simplicity as we focus on the argument $\bs_2$ here. From (\ref{fixedmean}) with $\bq_2=\boldsymbol{\mu\mu}^T$, we write $C(\bs_2)$ as
\begin{equation}
C(\bs_2)=\max_{\mathbf{y} \in \Omega(\bs_2)} f(\mathbf{y})
\end{equation}
where $f(\mathbf{y})=\frac{1}{2}\log |\mathbf{I}+\mathbf{H}\bq_1\bh^T|-\frac{1}{2}\log |\mathbf{I}+\bg\bq_1 \bg^T|$. Note that in this case $f(\mathbf{y})$ depends only on the first $t^2$ elements of $\mathbf{y}$. Now, we use \cite[Theorem 7]{pointsetmap}, which states conditions on the continuity of the optimal value function in mathematical programming to prove the continuity of $C(\bs_2)$. In the sequel, we verify that all requirements of \cite[Theorem 7]{pointsetmap} are satisfied.

Since the determinant of an $n \times n$ matrix $\mathbf{A}$ can be written as $\text{det}(\mathbf{A})=\sum_{\sigma} \text{sgn}(\sigma) \prod_{i=1}^n a_{i\sigma(i)}$, where the sum is over all $n!$ permutations of $\{1,2, \cdots, n\}$, the determinant in this form is a polynomial in $n^2$ variables, and $\det(\mathbf{A})$ is continuous. Consequently, $f(\mathbf{y})$ is also continuous. $\Omega(\bs_2)$ consists of linear matrix inequalities, hence it is a continuous point-to-set map. Furthermore, $\Omega(\bs_2)$ is uniformly compact  because for any sequence $\bs^{(i)}_2$ in the neighborhood of $\bs_2$, i.e., the metric distance $d(\bs^{(i)}_2,\bs_2)=\tr\left((\bs^{(i)}_2-\bs_2)(\bs^{(i)}_2-\bs_2)^T\right)\leq\delta^2$ for some finite $\delta>0$, one can find $k_i=\max \lambda(\bs^{(i)}_2)$ where $\lambda(\bs^{(i)}_2)$ is an eigenvalue of matrix $\bs^{(i)}_2$ such that
\begin{equation}
\Omega(\bs^{(i)}_2) \subseteq \mathcal{Y}=\{\text{row concatenation of }(\bq_1,\bq_2): \bq_1,\bq_2 \succeq \bo,\:  \tr(\bq_1+\bq_2) \leq k\}
\end{equation}
where $k=\max_i k_i \leq P+\delta$, where $P$ is the power constraint imposed on $\mathcal{S}_{PE}$. Since $\mathcal{Y}$ is compact and contains $\bigcup_{i} \Omega(\bs^{(i)}_2)$, $\Omega(\bs_2)$ is uniformly compact. Hence, the requirements of \cite[Theorem 7]{pointsetmap} are satisfied and $C(\bs_1,\bs_2,\bh,\bg)$ is continuous with respect to $\bs_2$.

\renewcommand{\baselinestretch}{1.4}
\bibliographystyle{IEEEtran}
\bibliography{IEEEabrv,myLibrary}

\end{document}